\documentclass[review]{elsarticle}

\usepackage{lineno,hyperref}
\usepackage{amsmath}
\usepackage{amssymb}
\usepackage{graphics}
\usepackage{graphicx}
\usepackage{epsfig}
\usepackage{dcolumn}
\usepackage{bm}
\usepackage{lineno}
\usepackage{multirow}
\usepackage{ragged2e}
\usepackage{rotating}
\usepackage{floatrow}
\usepackage{graphicx}
\newfloatcommand{capbtabbox}{table}[][\FBwidth]
\usepackage{blindtext}
\usepackage{caption}
\usepackage{wrapfig}
\usepackage{amsmath}
\usepackage{amssymb}
\usepackage{amsmath}
\usepackage{placeins}
\usepackage[english]{babel}
\usepackage{blindtext}

\begin{document}
	
	\begin{frontmatter}
		
		\title{J/$\psi$ yields in low energy nuclear collisions at SPS and FAIR: a baseline estimation}
		
		\author[add1]{S.~Chatterjee}
		\cortext[mycorrespondingauthor]{Corresponding author}
		
		\author[add2,add3]{P.~P.~Bhaduri\corref{mycorrespondingauthor}}
		\ead{partha.bhaduri@vecc.gov.in}
		
		\author[add2,add3]{S.~Chattopadhyay}
		
		\address[add1]{Department of Physics, Bose Institute, EN-80, Sector V, Kolkata-700091, India}
		\address[add2]{Variable Energy Cyclotron Centre, Bidhan Nagar, Kolkata-700064, India}
		\address[add3]{Homi Bhabha National Institute, Anushakti Nagar, Mumbai 400094, India}
		
		\begin{abstract}
			The yield of $J/\psi$ mesons, produced in proton-nucleus ($p+A$) and nucleus-nucleus ($A+A$) collisions are estimated within a Glauber model ansatz for the upcoming low energy heavy-ion collision experiments at SPS and FAIR. A data driven parametrization is employed to incorporate the effects of Cold Nuclear Matter (CNM) on the $J/\psi$ production cross-section.
		\end{abstract}
		
		
	\end{frontmatter}

	\section{Introduction} 
	
	The study of $J/\psi$ production in relativistic nuclear collisions, is considered as a promising diagnostic tool to characterize the hot and dense matter produced in the collisions, since the seminal work by Matsui $\&$ Satz in 1986~\cite{Matusi_staz}. The suppression of the production yield of $J/\psi$ mesons in heavy-ion collisions with respect to proton-proton~(p+p) and proton-nucleus~(p+A) collisions was proposed as an experimental signature of deconfinement transition and formation of Quark-Gluon Plasma (QGP)~\cite{Vogt}. Debye screening of the confining potential in a partonic medium was originally proposed as the only QGP effect causing dissociation of charmonium states. By now it is well known that suppression to a large extent is also induced by the medium-induced imaginary potential of the quarkonium system. In transport models of heavy-ion collisions, this is dubbed in terms of the inelastic collisions with hard gluons leading to the thermal broadening of the width of the in-medium quarkonium spectral function (see Ref.~\cite{Rothkopf:2019ipj} for a detailed review of the recent progress achieved in understanding quarkonium under extreme conditions, from a theoretical perspective.)
	
	Soon after the Matsui-Satz prediction, experimental investigations carried out at SPS
	identified a considerable amount of $J/\psi$ suppression already present in p+A collisions~\cite{NA60_158_400_GeV, NA50_400_GeV, NA50_450_GeV,NA50_400_GeV}, when compared with the yield in p+p collisions. A detailed survey on charmonium production at energies available at CERN-SPS and BNL-RHIC can be found in Ref.~\cite{survey_sps_RHIC}. In proton induced collisions, QGP or in general formation of any secondary medium is traditionally not anticipated. However, the results from the recent experimental programs at the LHC as well as at RHIC, have raised the question of whether there is a hydrodynamic medium created in these so-called smaller systems~\cite {smaller_system_1,smaller_system_2}. This change of paradigm may have consequences for lower energies, in particular at large charged particle multiplicity. As this article is focused on the low energy domain of the heavy-ion collisions, where the occurrence of the high multiplicity events would be extremely rare, hence it might be justified to assume the absence of any secondary medium in proton induced collisions. The p+p collisions might be considered to mimic the QCD vacuum. On the other hand, in p+A reactions, the nascent $J/\psi$ mesons during the pre-resonance or resonance stage of evolution may interact with the nucleons present in the target nucleus, the so-called primary medium, which might lead to their dissociation. Quantification of such Cold Nuclear Matter (CNM) effects were traditionally attempted within the Glauber model framework, by analyzing the target mass dependence of the production cross-section of $J/\psi$ mesons in p+A collisions~\cite{LABref}. An effective ``absorption'' cross-section $\sigma_{\rm J/\psi}^{eff}$ measures the overall suppression present in the data~\cite{NA60_158_400_GeV, NA50_400_GeV,NA50_450_GeV,NA50_400_GeV}. Systematic analysis of the data collected in these p+A reactions reveals a significant dependence of the absorption cross-section on collision energy with more dissociations at lower energies, as originally predicted in Ref.~\cite{lourenco_JHEP}. Experimental confirmation of this fact was subsequently performed by the NA60 Collaboration at SPS~\cite{arlandi}. The measurements revealed that in 158~GeV p+A collisions the $J/\psi$ absorption cross-section directly extracted from data is almost twice as large as that at 400 GeV. Estimation of the CNM suppression with high precision is an essential prerequisite to correctly interpreting the J/$\psi$ data collected in nucleus-nucleus (A+A) collisions and to isolate the genuine effects of a hot and dense secondary medium. 
	Extensive measurements of $J/\psi$ production in heavy-ion collisions were performed at SPS: in 158~A~GeV Pb-Pb collisions by the NA50 Collaboration~\cite{alessandro} and subsequently in 158 A~GeV In-In collisions by the NA60 Collaboration~\cite{arlandi_In_In}. The pertinent earlier significant experimental works at the SPS on the $J/\psi$ production can be found in Ref.~\cite{SPS_Jpsi,SPS_2,SPS_3,SPS_4,SPS_5,SPS_6,SPS_7,SPS_8,SPS_9,SPS_10,SPS_11,SPS_12}. Within errors, the relative $J/\psi$ yield in In-In collisions has been found to be in line with suppression induced by cold nuclear matter; an additional suppression of about 25 - 30 $\%$ is observed in the most central Pb-Pb collisions. The conceptual origin of this anomalous suppression is not yet clear, a variety of models with and without plasma suppression effects are available in the literature, none of which can precisely explain the data. The reader may note that the statistical regeneration picture of charmonium production, a feature found essential at RHIC and LHC energies, was also applied to the SPS heavy-ion data~\cite{PBM}. 
	
	In nuclear collisions, no measurement on $J/\psi$ production below top SPS energy (158 A GeV) has been attempted till date, owing to extremely low charm production cross-sections. $J/\psi$ measurements in low energy collisions demand accelerators with unprecedented beam intensities and detectors with very high rate capabilities. The Compressed Baryonic Matter (CBM) experiment, currently being constructed at the Facility for Anti-proton and Ion Research (FAIR) accelerator complex in Darmstadt, Germany aims at the measurement of $J/\psi$ mesons via their decay into dileptons (both e$^{+}$e$^{-}$ and $\mu^{+}\mu^{-}$ channels) in proton and ion induced collisions~\cite{CBM_Physics_book}. The SIS100 accelerator at FAIR will deliver accelerated proton beams with beam energy up to 30 GeV, while for heavy ions~(Z/A~$\sim$~0.4) the foreseen beam energy ranges between 3.43~-~12.04~A~GeV and for light ions~(Z/A~$\sim$~0.5) 3.43~-~15.0~A GeV~\cite{i_augustin,p_senger}, corresponding to centre-of-mass energy $\sqrt{s_{NN}}$~$\sim$~2.7~-~5.5~GeV. The foreseen unprecedented beam intensities for both protons and heavy-ions at FAIR would enable the detection of $J/\psi$ mesons in p+A and nucleus-nucleus (A+A) collisions. At FAIR, $J/\psi$ production would occur close to its kinematic production threshold ($E_{b}^{th}~\sim12.5$~GeV for an elementary $NN$ reaction like: $N +N \rightarrow N+N+J/\psi$). The physics perspectives of the planned $J/\psi$ measurements~\cite{t_ablyazimov} include a stringent test of the perturbative QCD (pQCD) based models of charmonia production near threshold, the investigation of charm production and propagation through dense baryonic medium and the possibility of discovering the predicted sub-threshold charm production~\cite{j_steinheimer} in ion-ion collisions among the others.
	
	Another complementary experiment to CBM is the NA60+ experiment at CERN-SPS~\cite{na60+}. The NA60+ experiment plans to extend the existing study on charmonia production by performing the $J/\psi$ measurement via the di-muon decay channel in the beam energy range 40 A to 160 A GeV, corresponding to the center of mass energy range of $\sqrt{s_{NN}}$~$\sim$~6~-17~GeV. Thus with the data from CBM and NA60+, the dynamics of the $J/\psi$, likely to be produced in the early stages of the collisions, can be probed over a large energy range, in a rather unexplored energy domain. In addition to the unavailability of A+A data, no systematic measurements have been performed so far on charmonium production below 158 GeV in p+A collisions. This certainly calls for phenomenological investigations on $J/\psi$ production at low energies relevant for the upcoming experimental programs. A suitable estimation of $J/\psi$ yields in this energy domain would be extremely useful to optimize and reliably access the physics capabilities of charmonium measurements of the corresponding experimental setups.
	
	In the present article, we adopt the conventional geometrical Glauber formalism for calculating the cross-section and yield of the $J/\psi$ mesons, produced in the early stages of p+A and A+A collisions at foreseen CBM and NA60+ energies. In literature, $J/\psi$ production in elementary collisions is generally modelled as a two step process: initially compact $c\bar{c}$ pairs produced via hard scattering are treated with perturbative QCD (pQCD), whereas subsequent resonance binding of the pair is non-perturbative in nature and treated phenomenologically. The application of such two component models is based on QCD factorization, which separates the initial $c\bar{c}$ production from the formation of the bound state. However at low collision energies, like those will be available at FAIR or CERN-SPS, the validity of the QCD factorization is questionable. We thus employ here a theory agnostic data driven parametrization to calculate the $J/\psi$ production in elementary $p+p$ collisions. The corresponding production cross-section in $p+A$ and $A+A$ collisions are estimated employing the Glauber framework, where the so called CNM suppression effects are incorporated via an effective absorption cross-section. No additional medium effect is incorporated in our calculations, even for A+A collisions. Our estimations would thus serve as a reference baseline for the upcoming measurements, with respect to which possible genuine secondary medium effects on charmonium production if present, can be isolated.
	
	The article is organized in the following way. In section 2, for completeness, we present a brief introduction to the Glauber model. $J/\psi$ production in elementary proton-nucleon (p+N) collisions is discussed in section 3, while that in p+A and A+A collisions within the present geometrical approach are provided in section 4. Section 5 presents the projected $J/\psi$ yield for different collision systems at different energies, relevant for the upcoming experimental facilities. We summarise our results in section 6.
	
	\section{Brief introduction to the Glauber model}
	In relativistic heavy-ion collisions, the Glauber model is common for a quantitative description of the geometrical configuration of the colliding systems~\cite{r_glauber} and related particle production. The model is based on the mean free path concept with a minimal set of assumptions like the baryon-baryon interaction cross-section remains unaltered throughout the passage of baryons of one nucleus into another and the nucleons follow a straight-line trajectory along the axis of collision. For two nuclei colliding at a certain impact parameter, the model is useful to compute the number of binary nucleon-nucleon (NN) collisions and the number of participating nucleons among the others. 
	The Glauber model can be computed following two different approaches: the Optical approach which assumes a smooth distribution of the nucleons inside the colliding nuclei and the Monte Carlo (MC) approach which accounts for the fluctuating positions of the nucleons inside the nucleus. In this article, the optical approach is employed to calculate the $J/\psi$ production cross-sections at CBM and NA60+ energies. The basic assumptions behind the Optical approach are the following: a) colliding nucleons traverse via straight-line trajectories b) the nucleons move independently inside the nucleus c) the nuclear size is much larger than the spatial range of the nucleon-nucleon force. A detailed review of the Glauber model can be found in Ref.~\cite{l_millar}.  
	
	\section{$J/\psi$ production in p+N collisions}
	
	The hadroproduction of $J/\psi$ is studied over many years for different collision systems (e.g. p+p, p+$\bar{p}$, p+$\pi$ etc). As we are interested to study the $J/\psi$ production in p+A and A+A collisions, we will mainly consider the p+p systems as a reference for hadronic collisions. In all the experiments, $J/\psi$ is detected via the dileptonic decay channel, using the conventional technique of invariant mass reconstruction from the decay products. Since the data are available from experiments of different periods spanning around 40 years or more, therefore, it is very important to treat all the available data on an equal footing. In Ref.~\cite{Zha}, the authors have systematically analyzed the world data on inclusive $J/\psi$ production cross-section in proton induced interactions from CERN-PS to LHC energies ($\sqrt{s} = 6.8 - 7000$ GeV). For a consistent comparison of different experimental results, all the published values have been updated with the latest best known values branching ratios and kinematics. The functional forms used to account for the x$_{F}$ and p$_{T}$ distributions are respectively, $\frac{d\sigma}{dx_{F}}~=~a\times~e^{-ln2(\frac{x_{F}}{b})^{c}}$ and $\frac{d\sigma}{dp_{T}}~=~d\times~\frac{p_{T}}{(1+e^{2}p_{T}^{2})^f}$ where a, b, c, d, e, and f are free parameters. The branching ratio value used for the correction is 5.961~$\underline{+}$~0.032\%~for~$J/\psi~\rightarrow~\mu^{+}\mu^{-}$ and 5.971~$\underline{+}$~0.032\%~for~$J/\psi~\rightarrow~e^{+}e^{-}$~\cite{branching_ratio}. The $\sqrt{s}$ dependence of the per nucleon inclusive $J/\psi$ production cross section is fitted with a parametric function, which nicely describes the energy evolution over a very broad range. For our study, we adopt the same parametrization but independently fitted the inclusive $J/\psi$ production cross-section in p+p~\cite{pp_data_1,pp_data_2,pp_data_3,pp_data_4,pp_data_5,pp_data_6,pp_data_7}, p+Be~\cite{pBe_data_1,pBe_data_2,pBe_data_3,pBe_data_4} and p+Li~\cite{pLi} collisions, as reported in TABLE I of Ref.~\cite{Zha}. As we aim to include the cold nuclear effects explicitly via Glauber ansatz, we refrain to use the data points for proton induced collisions with nuclear targets heavier than Beryllium~(Be). Light nuclei like Lithium (Li) and Beryllium (Be) are believed to have negligible nuclear effects on charm production and were previously used as reference by different experiments~\cite{NA60_158_400_GeV}.

	
	\section{Parameterization of $J/\psi$ production cross-section}
		
\begin{figure*}[h!]
		\centering
		\includegraphics[width=8.5cm, height=7.0cm]{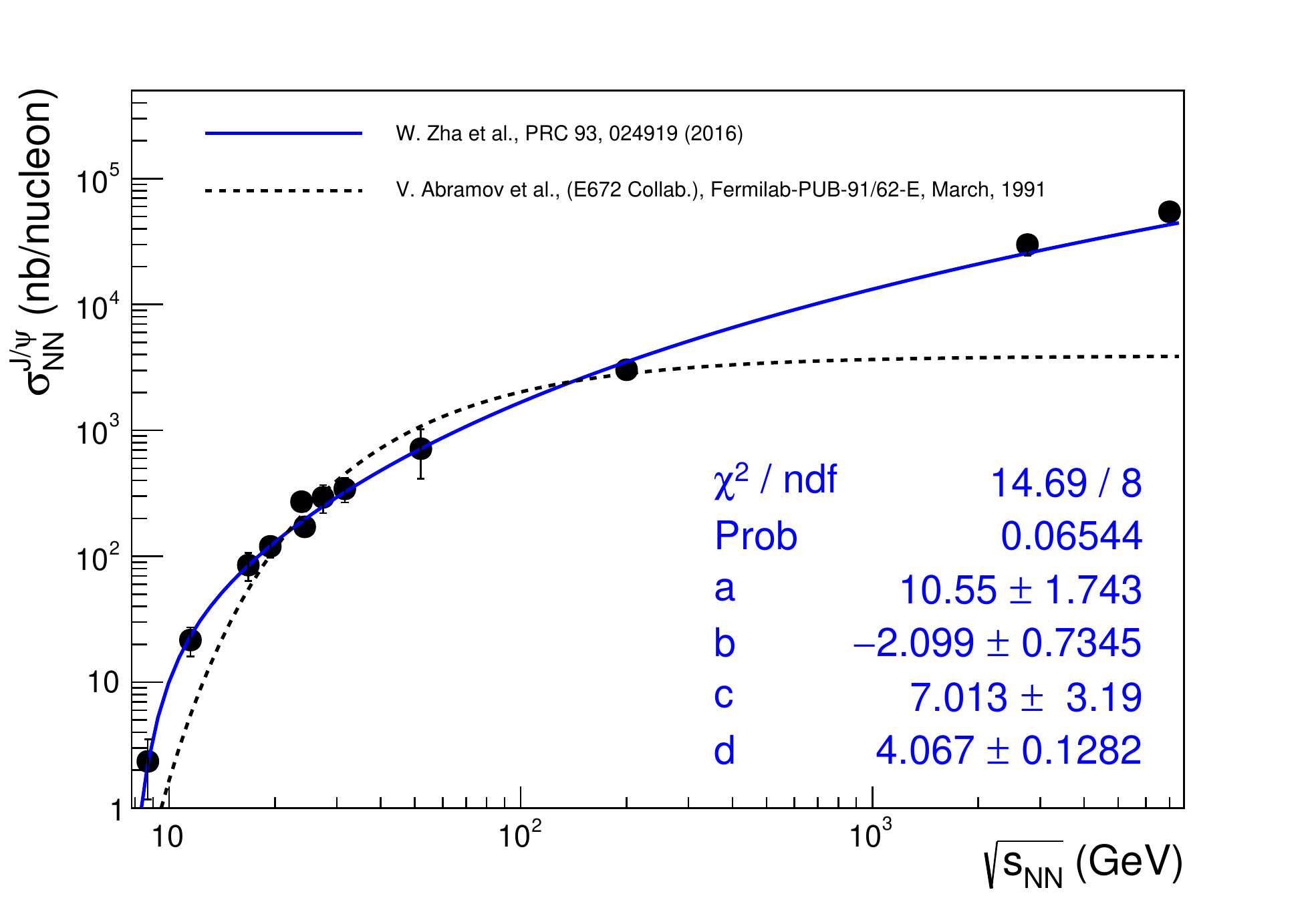}
		\caption{Energy dependence of inclusive $J/\psi$ production cross-section in p+p, p+Li and p+Be collisions.}
		\label{fig2}
	\end{figure*}

	The energy evolution of the $\sigma_{NN}^{J/\psi}$ in our adopted parameterization~\cite{Zha} is given below
	
	\begin{eqnarray}
	\label{eqn1}
	\sigma_{NN}^{J/\psi}(\sqrt{s}) &=& a \times y_{max}^{d} \times e^{\frac{-b}{y_{max}^{c}}}  
	\end{eqnarray} 
	where $y_{max} = ln(\frac{\sqrt{s}}{m_{J/\psi}})$, is the maximum value of the rapidity with which $J/\psi$ can be produced for a given centre-of-mass energy ($\sqrt{s}$). The free parameters a, b, c and d are fixed from fitting the selective p+N data subset as stated above. The MINUIT fit package based on minimization of $\chi^2$, available within ROOT~\cite{root} software framework is used in the present work. The variation of $\sigma_{NN}^{J/\psi}$ as a function $\sqrt{s}$ and its parameterization using Eqn.~\ref{eqn1} are shown in Fig.~\ref{fig2}, with a normalized $\chi{2}$ of 1.84 and fit $p$-value of 0.06. 
	A study of the covariant matrix associated with the fit reveals that all the four fit parameters are correlated. Other parameterizations are also available in the literature like the so called  Schuler parameterization obtained by the E672 collaboration~\cite{schuler}, the parameterization available from the E771 collaboration~\cite{E771} and also one given by Lourenco {\it et al.} in Ref.~\cite{lorenco}. These parameterizations use the same functional form for $\sigma_{NN}^{J/\psi}(\sqrt{s})$, but differ only in the value of the fitted parameters. However, none of these previous parameterizations is seen to explain the global data, particularly at higher energies. This can be seen from Fig.~\ref{fig2}. The dotted line represents the Schuler parameterization, with parameters fitted explicitly using our selected data corpus. The same holds for the other two parameterizations, though not explicitly shown here. Hence in this work, we use the parametrization given by Eqn.~\ref{eqn1} and extrapolate to predict the inclusive $J/\psi$ production cross-section at lower energies. The extrapolated values of $\sigma_{NN}^{J/\psi}$ using Eqn.~\ref{eqn1} at energies relevant for SPS and FAIR are given in Table~\ref{table2}.
	
	\begin{table} [h!]
		{
			\begin{tabular}{ |c|c|}
				\hline
				{\bf $\sqrt{s_{NN}}$} & {\bf {present work}} \\
				{\bf (GeV)}&{\bf $\sigma_{J/\psi}$(nb/nucleon)}\\
				\hline
				
				7.6 & 0.08 $\underline{+}$~0.12\\
				\hline
				9.8 & 8.61 $\underline{+}$~6.13 \\
				\hline
				12.3 & 31.30 $\underline{+}$~15.33 \\
				\hline
				17.0 & 87.44 $\underline{+}$~27.38 \\
				\hline
				
			\end{tabular}
		}
		\caption{The extrapolated inclusive production cross-sections of $J/\psi$~($\sigma_{J/\psi}$) mesons in p+p collisions at CBM and NA60+ energies using Eqn.~\ref{eqn1}. The errors correspond to the uncertainties associated with the four fitting parameters.}    
		\label{table2}
	\end{table}
	
	One may note the large error in the extrapolated $\sigma_{J/\psi}$ values, particularly at the lowest energy. The source of the errors is exclusively the uncertainty associated with the four fit parameters, correlated among themselves. 

	
	
	\section{$J/\psi$ production in $p+A$ and $A+A$ collisions}
	
	After discussing $p+p$ collisions, we move on to the measurements of proton beams incident on nuclear targets. In such $p+A$ collisions, $J/\psi$ production is influenced by the presence of the target nucleons. In experimental analysis, two approaches are in common to quantify the effect of nuclear matter on the charmonium ($J/\psi$) production in p+A collisions. The first one, the so-called $\alpha$-parametrization is generally followed to study the nuclear target dependence of all the hard processes including charmonium production. In this method the target mass (A) dependence of the charmonium production cross-section is fitted with a simple power-law \cite{wong}, $\sigma_{pA}^{J/\psi}= \sigma_{pp}^{J/\psi}\times A^{\alpha}$. $\sigma_{pA}^{J/\psi}$ and $\sigma_{pp}^{J/\psi}$ respectively denote the production cross-section of $J/\psi$ mesons in p+A and p+p collisions, and the exponent $\alpha$ is extracted from the data fitting, that encodes the nuclear matter effects. In absence of any cold matter effect, $\alpha$ is expected to be unity, whereas $\alpha~\textless 1$ indicates suppression. Due to the unavailability of $\sigma_{pp}^{J/\psi}$, different experiments have also used the lightest available target, to extract $\alpha$ from heavy-to-light target ratio of production cross section. The E866 experiment at FNAL, with 800 GeV proton beam, obtained a $\alpha$ value around 0.95 near mid rapidity ($x_{F}=0$)~\cite{E866}. The NA50 experiment at CERN-SPS in the similar kinematic range reported $\alpha= 0.925 \pm 0.015$ at 450 GeV and $\alpha = 0.925 \pm 0.005$ at 400~GeV~\cite{NA50_400_GeV,NA50_450_GeV}. The following NA60 experiment at CERN-SPS have found $\alpha = 0.927 \pm 0.013  (stat) \pm 0.009 (syst)$ at 400 GeV and $\alpha = 0.882 \pm 0.009 (stat) \pm 0.008 (syst)$ at 158 GeV~\cite{NA60_158_400_GeV}. The results indicate amplification of the overall cold matter suppression at lower collision energies. A detailed comparison of the $x_F$ dependence of the $\alpha$ parameter, obtained from different fixed target p+A collisions, at varying collision energies revealed two main features. At a given $x_F$ (or rapidity), $\alpha$ decreases with the reduction in the energy of the incident proton beam, as was seen in the inclusive analysis. $\alpha$ is also seen to decrease steadily with an increase in $x_F$, indicating more suppression at forward rapidity. A satisfactory theoretical interpretation of the observed $\alpha(x_{F})$ pattern is yet to be available. Various models have incorporated different cold matter effects, operative at different stages of production~\cite{model_1,model_2,model_3,r_vogt_model}. These include the final state dissociation, initial state modification of parton densities inside the target nucleus, energy loss in the initial and final stage and contribution of an intrinsic charm component in the nucleon wave function. However relative importance of these different effects is still unsettled. On passing, it is also worth mentioning that $\alpha$ as extracted from the data is sensitive to the chosen lightest target nucleus. Experiments that compare heavy targets with Hydrogen or Deuterium systematically derive artificially high values of $\alpha$~\cite{ref_1}. For example, the $\alpha$ values obtained from the measured $J/\psi$ production cross section ratios between p+p and p+Pt collisions by the NA3 experiment at 200 GeV are in partial contradiction with those results which use beryllium as the lightest target and was corrected for a meaningful comparison~\cite{ref_2}.
	
	The formalism widely used at SPS to analyze charmonium production in p+A collisions is based on the Glauber model calculations. As stated above, the model treats a p+A or A+A interaction as a collection of independent interactions between the target and projectile nucleons. The properties of the nucleons are assumed to remain unchanged after the first interaction and they can interact further with the same cross section. Following this approach, the $J/\psi$ production cross-section in p+A collisions can be expressed as \cite{wong}; 
	\begin{equation}
	\label{eqn3}
	{\sigma_{pA}^{J/\psi}} = {\frac{\sigma_{pp}^{J/\psi}}{\sigma_{abs}^{J/\psi}}} \int db_{A} [1-(1-T_{A}(b_A)\sigma_{abs}^{J/\psi})^A]
	\end{equation}  
	where $\sigma_{abs}^{J/\psi}$ denotes the $J/\psi$ absorption cross-section, accounting for the dissociation cross-section of the evolving c$\bar{c}$ pairs in the pre-resonance or resonance stage because of their inelastic interaction with the target nucleons along the path. $T_A(b_A)$ is the target nuclear thickness function, the basic input of the formalism for p+A collisions and represents the nucleons per unit surface at the impact parameter $b_{A}$. For large values of A, one can approximate Equation \eqref{eqn3}, in the first order, by an exponential parametrization as 
	
	\begin{equation}
	\label{eqn4}
	{\frac{\sigma_{pA}^{J/\psi}}{A}} = {\sigma_{pp}^{J/\psi}} exp(-\sigma_{abs}^{J/\psi}\rho L)
	\end{equation}    
	This is popularly called "$\rho L$ parameterization", where $\rho L$ represents the average amount of nuclear matter traversed by the nascent charmonium state from the point of production inside the target nucleus till it exits the nuclear medium. 
	
	It should be noted that $\sigma_{abs}^{J/\psi}$ as implemented in the above two equations and extracted from the fitting of the measured $J/\psi$ cross-sections are effective quantities encoding the convolution of all effects of the cold nuclear medium responsible for the reduction of $J/\psi$ yield. Within such purely data driven approaches, it is not possible to disentangle the magnitude of different effects leading to an overall reduction.
	
	\subsection{\textbf {Extraction of $J/\psi$ absorption cross-section}}
	
	Explicit analysis of SPS data has shown that the fit results corresponding to  "$\rho L$ parameterization" are almost indistinguishable from full Glauber calculations~\cite{NA50_400_GeV}. Hence in our present work, for simplicity, we have opted the $\rho L$ parametrization, with $\rho =\rho_{0}$, the saturation nuclear density. 
	$L(A)$ is calculated following prescription given in~\cite{wong}, assuming a spherical nucleus of uniform density. We analyse the data available on inclusive $J/\psi$ production cross-sections in $p+A$ reactions, as available from different fixed-target experiments at SPS. The data corpus used is summarised in Table~\ref{table3}. 
		\begin{table} [h!]
		\resizebox{\columnwidth}{!}
		{
			\begin{tabular}{ |c|c|c|c|}
				\hline
				{\bf Experiment} & {\bf $\sqrt{s_{NN}}$ (GeV)} & {\bf targets } & {\bf phase space}   \\
				\hline
				NA60~\cite{NA60_158_400_GeV}& 17.27  & Be, Al, Cu, In, W, Pb, U & 0.28$<$y$_{cms}$$<$0.78; 0.05$<$x$_{F}$$<$0.40\\
				\hline
				NA60~\cite{NA60_158_400_GeV}& 27.42 & Be, Cu, Ag, W, Pb,U  & -0.17$<$y$_{cms}$$<$0.575; -0.075$<$x$_{F}$$<$0.125\\
				\hline
				NA50~\cite{NA50_400_GeV}& 27.42 & Be, Al, Cu, Ag, W, Pb & -0.425$<$y$_{cms}$$<$0.33\\
				\hline
				NA50~(HI)~\cite{NA50_400_GeV}  & 29.08 & Be, Al, Cu, Ag, W & -0.425$<$y$_{cms}$$<$0.33; -0.1$<$x$_{F}$$<$0.1\\
				\hline
				NA50~(LI)~\cite{NA50_400_GeV}  & 29.08 & Be, Al, Cu, Ag, W & -0.425$<$y$_{cms}$$<$0.33; -0.1$<$x$_{F}$$<$0.1\\
				\hline        
			\end{tabular}
		}
		\caption{ Summary of the data taken from the experiments at SPS for the p+A collisions and their phase space coverage.}    
		\label{table3}
	\end{table}
	The NA50 experiment measured charmonium production in p+A collisions at 400 GeV and 450 GeV~\cite{NA50_400_GeV,NA50_450_GeV}. At 450 GeV, five different nuclear targets were exposed to the beam. For each target two qualitatively different data samples were collected, with varying intensity of the incident proton beam. 
	Absolute values of the per nucleon inclusive production cross sections in the dimuon channel over the kinematic domain $-0.5 < y_{cms} < 0.5$ and $-0.5 <cos(\theta)_{cs} < 0.5$ were published separately for HI and LI data samples for each target nucleus. The extracted per nucleon cross sections from the LI data samples are systematically higher than HI values, the relative difference being $5 \%$. The dominant contribution to the systematic errors was attributed to the uncertainties in the factors required for calculating the absolute normalization: luminosity, trigger efficiency and dimuon reconstruction efficiency. 
	It may be noted here, that for a coherent comparison, the 450~GeV results were corrected so that they correspond to the same analysis conditions used for the 400 GeV data sample. We use this reanalyzed latest results reported in Ref.~\cite{NA50_400_GeV} for 450~GeV beam energy in our analysis. In addition to absolute cross sections data were also published in terms of the $J/\psi$-to-Drell-Yan ($\psi$-to-DY) ratio, where dominant contribution to the uncertainties came from the statistical errors of the DY data sample~\footnote{The DY muon pairs were collected in the mass range 2.9~$<$~m$_{\mu\mu}<$~4.5~GeV/c$^2$. The kinematic coverage was same as that of the $J/\psi$.}. The major motivation behind the p+A data taking by the NA60 experiment was to look for a precise and robust benchmark for interpreting the heavy-ion results. 
	Data samples were collected in the rapidity region $0.28 < y_{cms} < 0.78$ at 158 GeV and $-0.17 < y_{cms} < 0.33$ at 400 GeV. Results were published in terms of relative production cross sections, where per nucleon absolute production cross section for each target nucleus with mass number $A$ was normalized to the cross section for the lightest target (Be). 
	It is important to mention that we have not included in our study, the p+A collisions with 200 GeV proton beam for a variety of nuclear targets, performed by the NA38 experiment at SPS. Subsequent analyses of this dataset made in the framework of the NA50 experiment~\footnote{The NA50 di-muon setup was based on the NA38 experimental setup that made it possible to reevaluate the reconstruction performance.} revealed the bias in these early measurements due to faulty determination of reconstruction efficiencies~\cite{na38_re,na38_re_1}.

	As the heavy-to-light target ratio of the $J/\psi$ production cross sections gives better control over the systematics, in our analysis we make the ratio fit to all the data sets summarized in Table~\ref{table3}. The uncertainties associated with each data point include statistical and target specific systematic errors added in quadrature. Note that we do not consider any polarization or any change of polarization as a function of the collision energy of the selected muon pairs~\footnote{In SPS p+A runs, $J/\psi$ polarization was investigated by the NA60 experiment by studying the angular distribution of the decay muons. Polarization parameters namely $\lambda$, $\mu$, $\nu$ were extracted from a fit to muon angular distribution. Preliminary results analyzed in the helicity reference frame indicated $\mu$ to be compatible with zero in the entire kinematic range, with slightly negative $\lambda$ values at low $p_T$, approaching to zero at high $p_T$. $\nu$ values were close to zero in the entire $p_T$ range~\cite{arnaldi_polarisation}. A detailed review of the measurement of the angular distribution of leptons from $J/\psi$'s produced inclusively in proton-nucleus collision at $\sqrt{s}$~=~41.6~GeV can be found in Ref.~\cite{HERA_angular_dist}}. The nuclear effects in the lightest target, i.e. Be, are negligible.

\begin{figure*}[h!]
		\includegraphics[width=5.50cm, height=5.0cm]{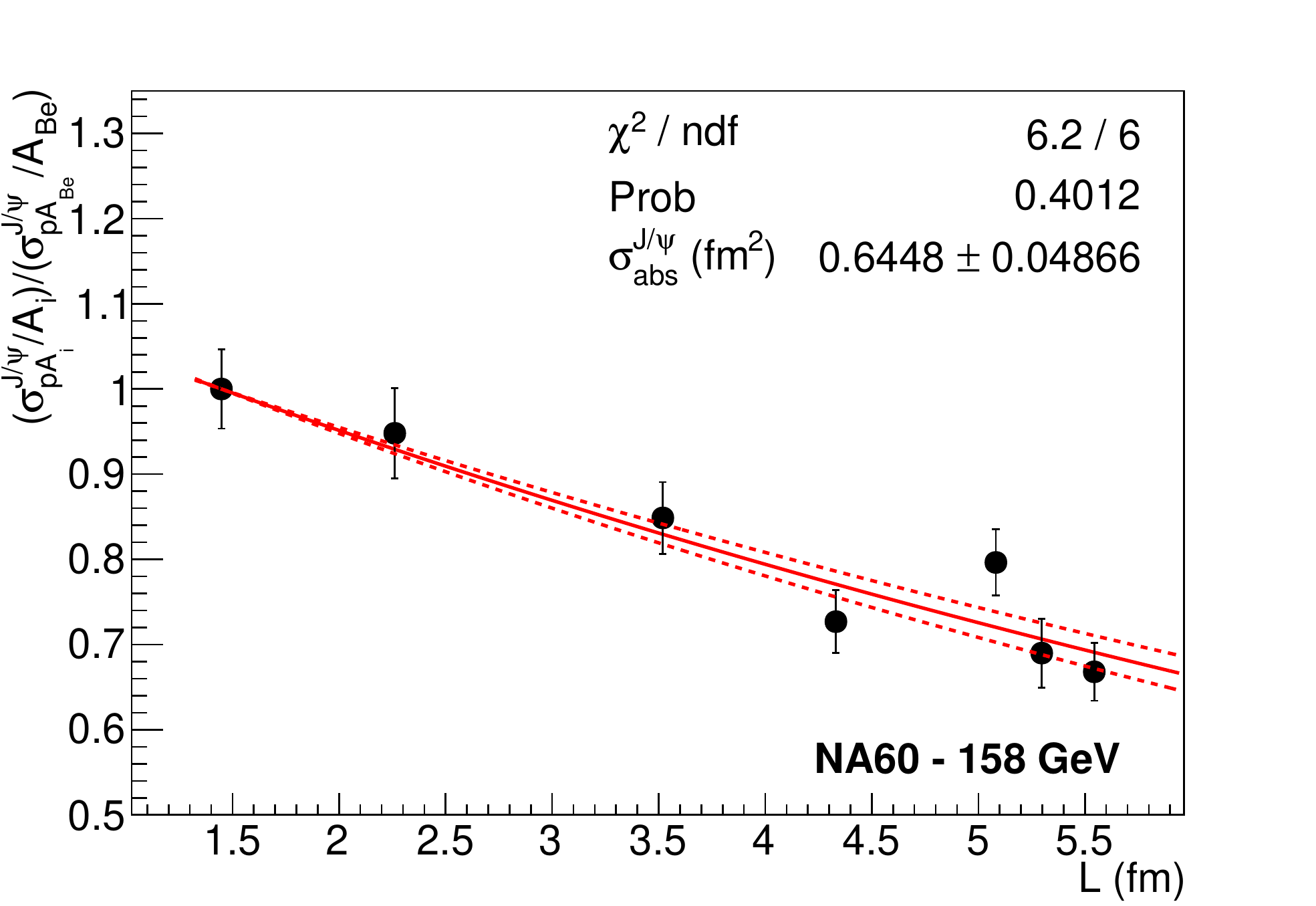}
		\includegraphics[width=5.50cm, height=5.0cm]{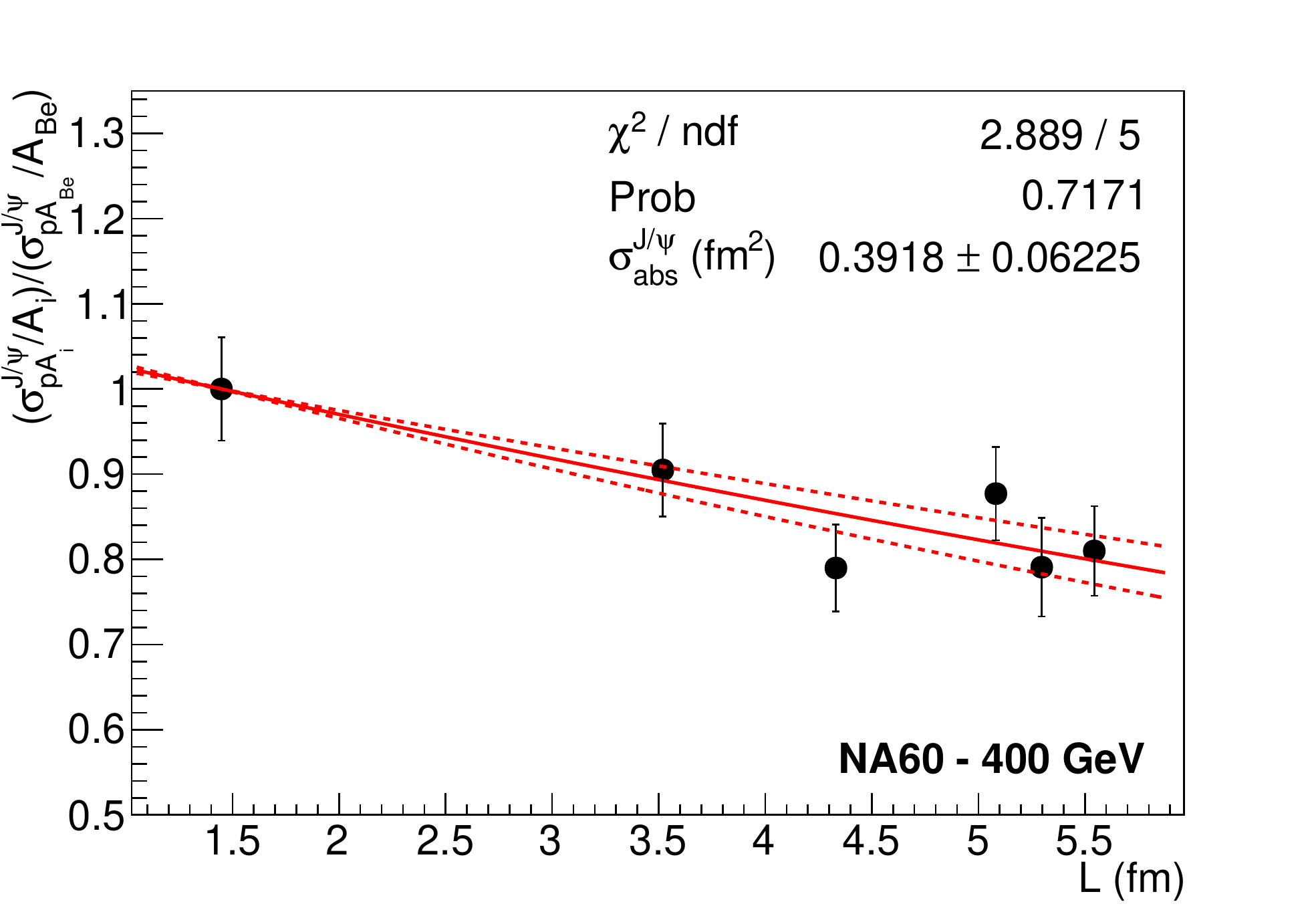}
		\includegraphics[width=5.50cm, height=5.0cm]{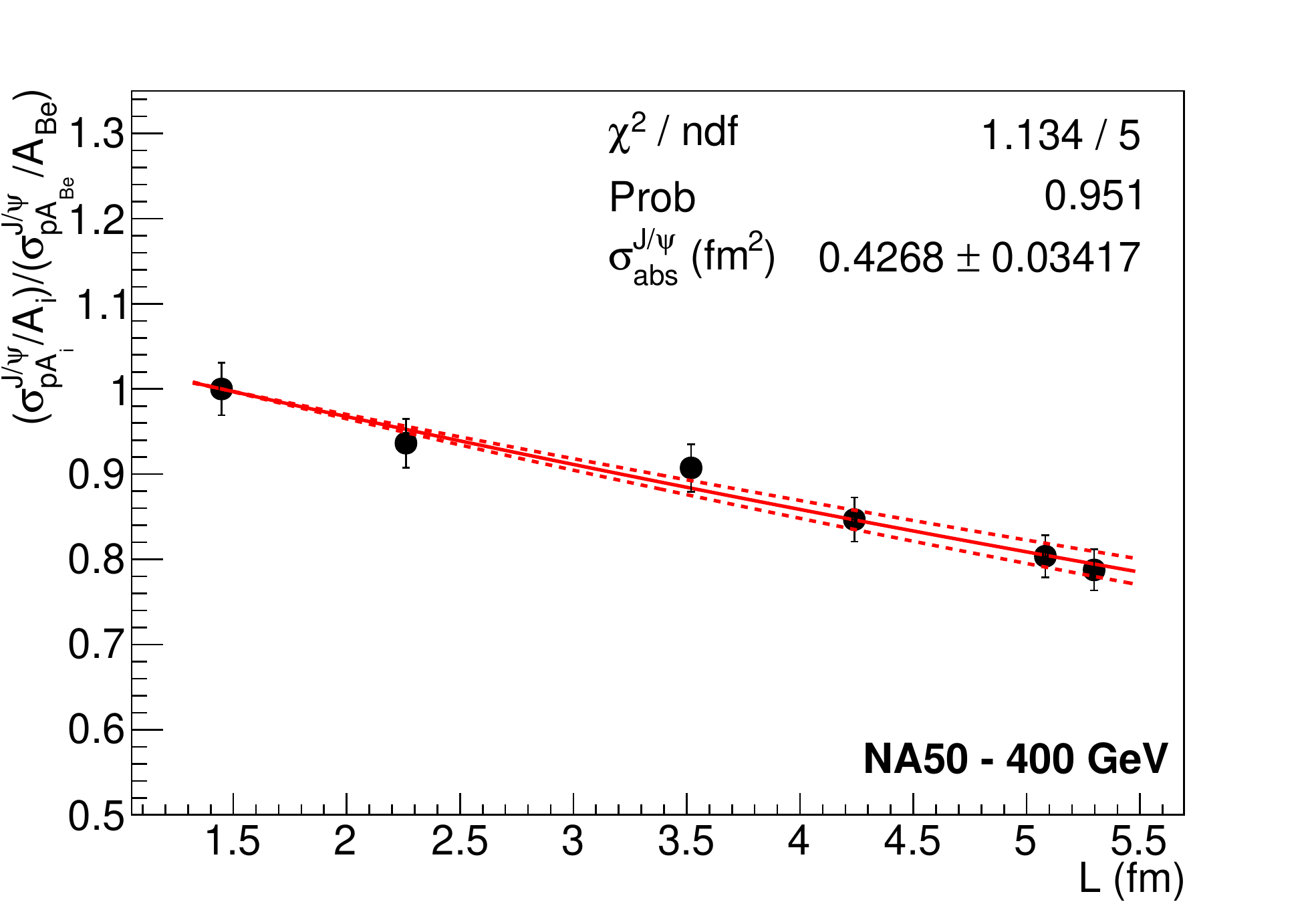}
		\includegraphics[width=5.50cm, height=5.0cm]{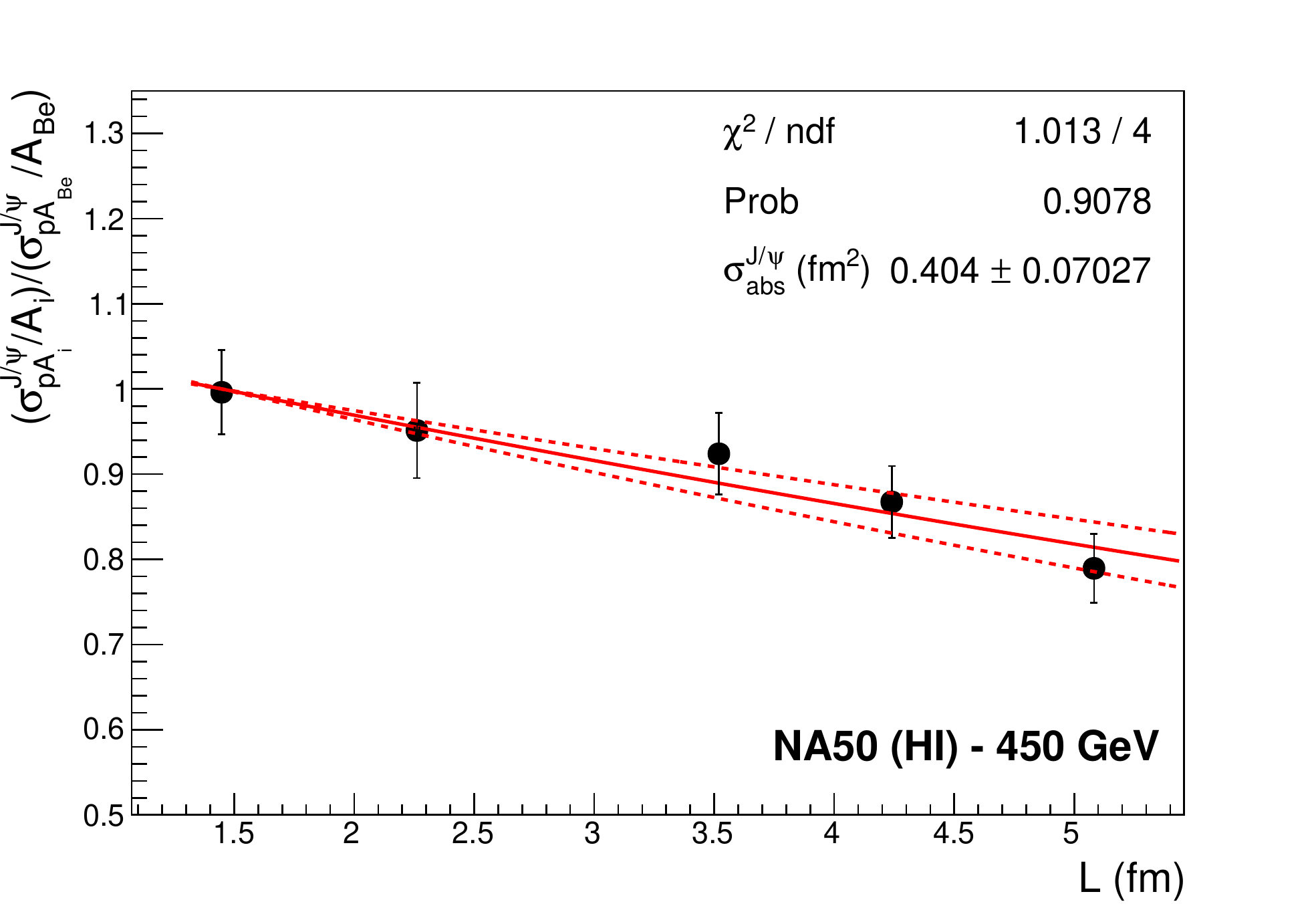}
		\includegraphics[width=5.50cm, height=5.0cm]{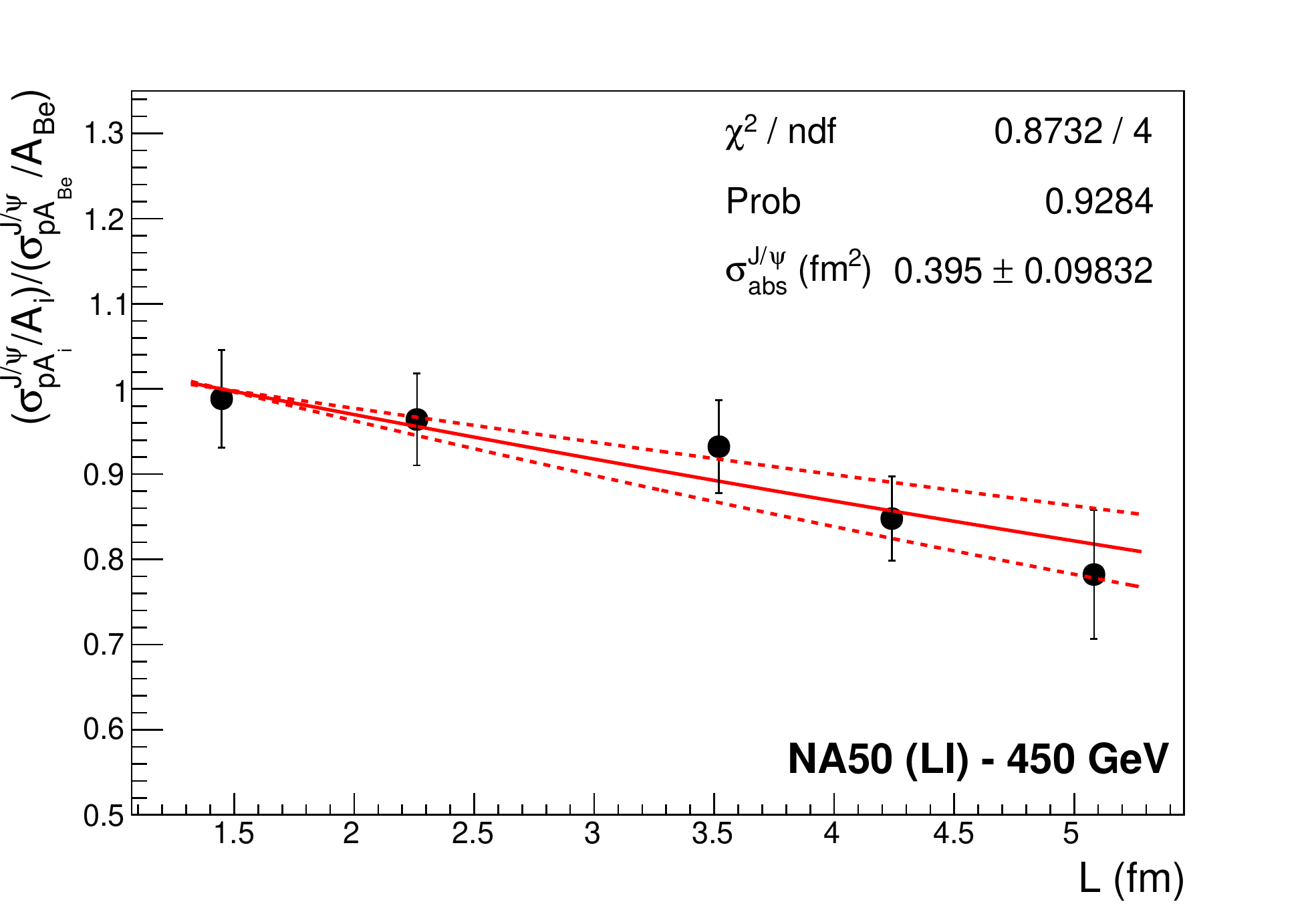}
		
		\caption{Variation of normalized $J/\psi$ production cross section as a function of the average distance~(L) of the nuclear matter of the target nucleus traversed by the $J/\psi$. Data are fitted following $\rho L$ parameterization as given in Eqn.~\ref{eqn4}. Error bars represent quadratically added statistical and target specific systematic uncertainties.}
		\label{fig4}
	\end{figure*}

	Our fit results using the $\rho L$ parametrization described by Eqn.~\ref{eqn4} are displayed in Fig.~\ref{fig4}. 
	The extracted values of the best fit parameter $\sigma_{abs}^{J/\psi}$ are summarised in Table~\ref{table4} along with the corresponding normalized $\chi^{2}$ and the $p$-value of the fit. The error values correspond to $1 \sigma$ fit uncertainty~(shown as the red dotted lines in Fig.~\ref{fig4}). Before we proceed further, it would be interesting to compare our extracted values of $\sigma_{abs}^{J/\psi}$ with existing results available in the literature. The NA50 Collaboration had estimated the $\sigma_{abs}^{J/\psi}$ both at 400 GeV and 450 GeV using both the full Glauber model and the $\rho L$ parametrization. For the most updated data sets, independent two parameter fits to the per nucleon absolute production cross-section, within full Glauber analysis results $\sigma_{abs}^{J/\psi} =  4.4 \pm 1.0$ mb for 450 GeV HI sample, $\sigma_{abs}^{J/\psi} =  4.1 \pm 1.4$ mb for 450 GeV LI sample and $\sigma_{abs}^{J/\psi} =  4.6 \pm 0.6$ mb at 400 GeV~\cite{NA50_400_GeV}. A simultaneous fit to all three above datasets, with a common absorption cross-section but two different normalizations (accounting for the difference in energy and rapidity coverage), leads to  $\sigma_{abs}^{J/\psi} = 4.5 \pm 0.5$ mb~\cite{NA50_400_GeV}. 
		\begin{table} [h!]
		
		\begin{tabular}{ |c|c|c|c|c|}
			\hline
			{\bf Experiment} & {\bf $\sqrt{s_{NN}}$ (GeV)}  & {\bf $\sigma_{abs}^{J/\psi}$ (mb)} & {\bf $\chi^{2}/ndf$} & Probability  \\
			\hline
			NA60& 17.27   & 6.45~\underline{+}~0.49& 1.03 & 0.40\\
			\hline
			NA60& 27.42  & 3.92~\underline{+}~0.62& 0.58& 0.72\\
			\hline
			NA50& 27.42 & 4.27~\underline{+}~0.34& 0.23& 0.95\\
			\hline
			NA50~(HI)  & 29.08 & 4.04~\underline{+}~0.70& 0.25& 0.91\\
			\hline
			NA50~(LI)  & 29.08  & 3.95~\underline{+}~0.98& 0.22& 0.93\\
			\hline        
		\end{tabular}
		\caption{ Value of $\sigma_{abs}^{J/\psi}$ extracted from the fitting of different experimental data on inclusive J/$\psi$ production in p+A collisions at SPS.}    
		\label{table4}
	\end{table}
	If instead of full Glauber, $\rho L$ parametrization is used, the corresponding $\sigma_{abs}^{J/\psi}$ values are found to be smaller by $10 \%$~\cite{NA50_400_GeV}. In addition to the absolute cross-section, the target mass dependence of the $\psi$-to-DY ratio was also studied by NA50 to evaluate the cold matter effects on charmonium production. Nuclear effects on Drell-Yan production in the initial or final states were assumed to be negligible. The extracted $\sigma_{abs}^{J/\psi}$ values in a full Glauber analysis comes out to be $\sigma_{abs}^{J/\psi} =  4.3 \pm 0.7$ mb for 450 GeV HI dataset, $\sigma_{abs}^{J/\psi} = 4.4 \pm 1.0$ mb for LI set at the same energy and $\sigma_{abs}^{J/\psi} = 3.4 \pm 1.2$ mb at 400 GeV~\cite{NA50_400_GeV}. A global fit to all three data sets gives $\sigma_{abs}^{J/\psi} = 4.2 \pm 0.5$ mb, a value closely agreeing with result extracted from a fit to absolute cross sections~\cite{NA50_400_GeV}. As discussed in Ref.~\cite{NA50_400_GeV},  NA50 400~GeV p+A data is particularly important in the determination of $\sigma_{abs}^{J/\psi}$ from a fit to the absolute production cross sections, due to the lower systematic uncertainties on the beam normalizations. On contrary, the measured data sets at 450 GeV corresponded to larger statistics and thus led to more accurate measurements of the $\psi$-to-DY  cross-section ratio, insensitive to beam normalization. The NA60 experiment evaluated the nuclear effects on $J/\psi$ production via the analysis of measured cross sections for each target, normalized to cross-section for the lightest one. Large statistical errors prevented the use of the $\psi$-to-DY ratio. Following Glauber analysis of the heavy-to-light cross section ratios, one obtains, $\sigma_{abs}^{J/\psi} = 7.6 \pm 0.7 (stat) \pm 0.6 (syst.)$ mb at 158 GeV and $\sigma_{abs}^{J/\psi} (400 GeV) = 4.3 \pm 0.8 (stat) \pm 0.6(syst)$ mb at 400 GeV~\cite{NA60_158_400_GeV}. Results indicate a stronger suppression at 158 GeV. On the other hand $\sigma_{abs}^{J/\psi}$ at 400~GeV is in good agreement with NA50 results at the same energy and comparable kinematic range. Since none of the above values of $\sigma_{abs}^{J/\psi}$ is corrected for initial state shadowing or any other cold matter effect, they represent an overall effective estimation of the nuclear effects on the $J/\psi$ production and can be directly compared to our results. As evident from Table~\ref{table4}, our estimated $\sigma_{abs}^{J/\psi}$ obtained from the ratio fit are in close agreement with the previous results at all energies. In this context, one may also note that in their Glauber analysis, to mimic nuclear density distribution NA50 used the Fermi Oscillator model for nuclei with $A < 17$ and Woods-Saxon parametrization for heavier ones. As mentioned before, in our analysis, we have used a constant density profile ($\rho =\rho_{0}$, nuclear saturation density) for all target nuclei, for estimating $L(A)$. The resulting effect on extracted $\sigma_{abs}^{J/\psi}$ value is seen to be negligible. 
		
	It might also be useful to compare our results with other phenomenological model calculations within the Glauber framework. In Ref.~\cite{Tram} the authors performed a global analysis of $J/\psi$ suppression in cold nuclear matter at SPS and RHIC. In elementary hadronic collisions, the $J/\psi$ production is determined within the leading order (LO) Color Evaporation Model (CEM). For nuclear targets, the initial state modifications of the target parton densities were accounted by a variety of nuclear Parton Distribution Functions (nPDF), and the final state absorption is incorporated via the Glauber model. A global fit to the available data collected with different beam particles and energies led to a single $\sigma_{abs}^{J/\psi}$ between 3 to 6 mb, the exact value of which was seen to be dependent on the choice of the PDF. The large systematic error in the extracted $\sigma_{abs}^{J/\psi}$ appeared due to the uncertainties in the nuclear gluon distributions. Prior to the publication of NA60 p+A results, in Ref.~\cite{lourenco_JHEP}, the authors employed Glauber formalism to analyze $J/\psi$ production cross sections in fixed target p+A collisions with incident proton beam energies varying between  200 to 920 GeV, and in d+Au collisions at RHIC, at $\sqrt{s_{NN}} = 200$ GeV. 
	
	\subsection{\textbf {Parameterization of $J/\psi$ absorption cross-section}}

\begin{figure*}[htb!]
		\includegraphics[width=0.60\textwidth]{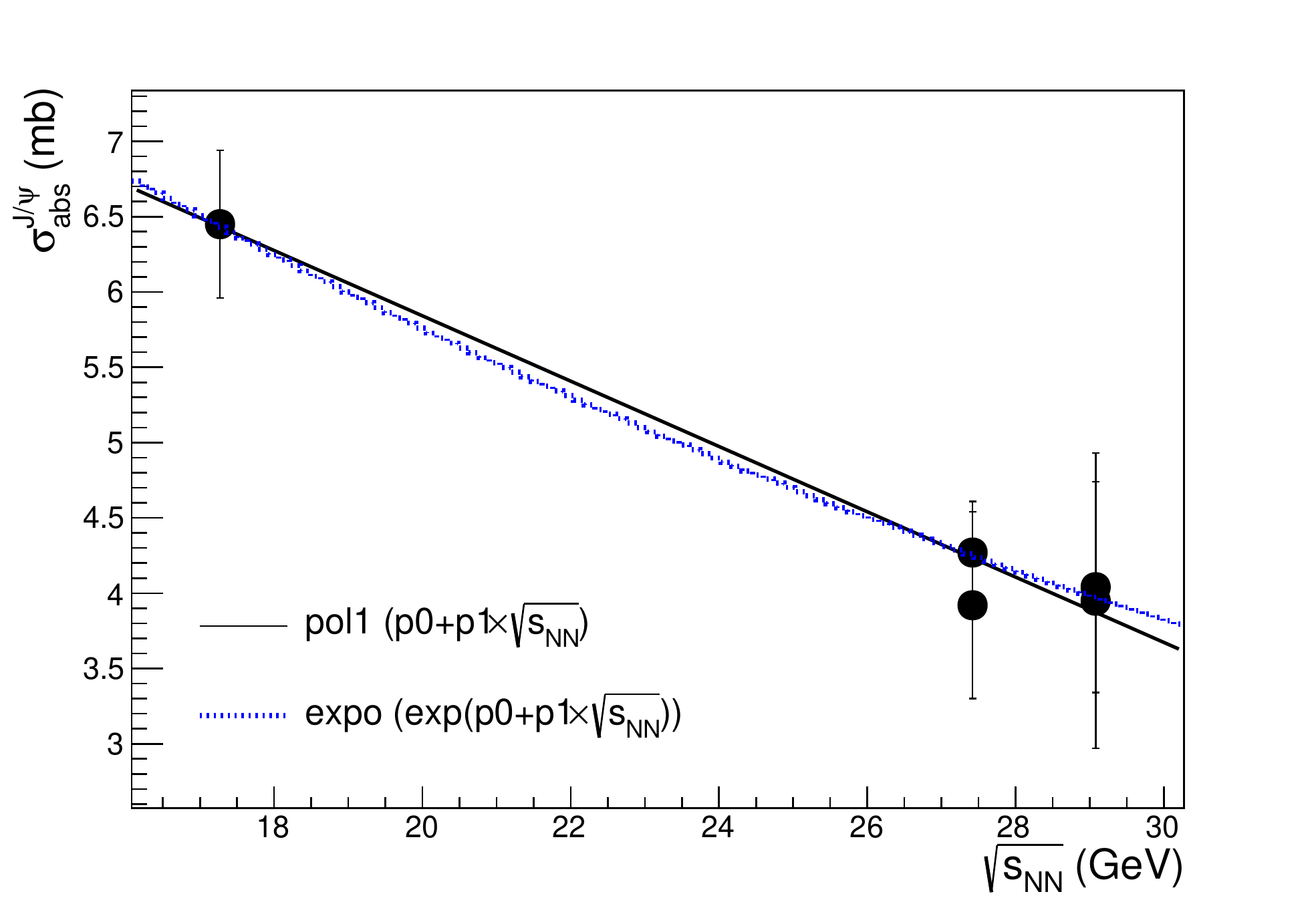}
		\caption{Variation of $\sigma_{abs}^{J/\psi}$ as a function of $\sqrt{s}$ and fitted with a 1$^{st}$ order polynomial and also with an exponential function.}
		\label{fig5}
	\end{figure*}

	Let us now proceed to estimate the magnitude of the CNM suppression at the foreseen energies of CBM and NA60+ experiments. For this purpose, we parameterize the centre-of-mass energy dependence of $\sigma_{abs}^{J/\psi}$ using two probable functions namely a  1$^{st}$ order polynomial of the form $\sigma_{abs}^{J/\psi}(\sqrt{s_{NN}})~=~p0+p1~\times~\sqrt{s_{NN}}$ and an exponential of the form $\sigma_{abs}^{J/\psi}(\sqrt{s_{NN}})~=~exp(p0+p1~\times~\sqrt{s_{NN}})$. For completeness, we have also attempted to parametrize the extracted $\sigma_{abs}^{J/\psi}$ values with an energy independent constant value. The resulting parameterizations are shown in Fig.~\ref{fig5} 
	and the fit results are summarized in Table~\ref{table_sigma_abs}.
	
	\begin{table} [h!]
		\resizebox{\columnwidth}{!}
		{
			
			\begin{tabular}{ |c|c|c|c|}
				\hline
				{\bf Fitting function} & {\bf Parameters} & {\bf $\chi^{2}/ndf$} & Probability  \\
				\hline
				pol0& p0~=~4.67~($\underline{+}$0.23)  & 4.35  & 0.002\\
				\hline
				pol1& p0~=~10.18~($\underline{+}$1.35), p1~=~-0.22~($\underline{+}$0.05)  &  0.11& 0.95\\
				\hline
				expo& p0~=~2.59~($\underline{+}$0.23), p1~=~-0.04~($\underline{+}$0.01)&  0.09 & 0.96\\
				\hline        
			\end{tabular}
		}
		\caption{ Parameterization of $\sigma_{abs}^{J/\psi}$ as a function of $\sqrt{s_{NN}}$. The details of the fit results are summarised in this Table.}    
		\label{table_sigma_abs}
	\end{table}
		
	The energy independent absorption scenario appears to be unlikely as indicated by the large $\chi^{2}$ and extremely small fit probability. Both energy dependent parameterizations have comparable~(within error bars) fit qualities. 
	\subsection{\textbf {Estimation of $J/\psi$ production cross-section in p+A and A+A collisions}}
	The values of $\sigma_{abs}^{J/\psi}$ for CBM and NA60+ energies as extrapolated from the linear parameterization~(pol1) along with the corresponding values of $J/\psi$ production cross-section using $\rho L$ parametrization, for a variety of nuclear targets are listed in Table~\ref{table6}.	As our analysis so far is based on rapidity integrated cross sections around central rapidity, hence the quoted numbers should be interpreted as the inclusive cross-sections around mid-rapidity.  	
	
	\begin{table} [h!]
		{
			\begin{tabular}{ |c|c|c|c|}
				\hline
				{\bf $\sqrt{s_{NN}}$}&7.6 GeV&12.3 GeV&17.0 GeV\\
				\hline 
				{\bf $\sigma_{abs}^{J/\psi}$ } (mb)&8.51$\underline{+}$ 1.40&7.47 $\underline{+}$ 1.48&6.44 $\underline{+}$ 1.60\\
				\hline 
				{\bf $\sigma_{pAl}$} (nb/nucleon)&0.06 $\underline{+}$ 0.09 &24.71 $\underline{+}$ 13.26 &71.31 $\underline{+}$ 25.94\\
				\hline 
				{\bf $\sigma_{pNi}$} (nb/nucleon)&0.05 $\underline{+}$ 0.08 & 21.84 $\underline{+}$ 12.26 & 64.12 $\underline{+}$ 25.02\\
				\hline
				{\bf $\sigma_{pCu}$} (nb/nucleon)&0.05 $\underline{+}$ 0.08 & 21.66 $\underline{+}$ 12.19 & 63.66 $\underline{+}$ 24.95\\
				\hline 
				{\bf $\sigma_{pW}$} (nb/nucleon)&0.04 $\underline{+}$ 0.07 &18.40 $\underline{+}$ 10.95 &55.30 $\underline{+}$ 23.61\\
				\hline
				{\bf $\sigma_{pAu}$} (nb/nucleon)&0.04 $\underline{+}$ 0.07&18.17 $\underline{+}$ 10.86 &54.71 $\underline{+}$ 23.51\\
				\hline
				{\bf $\sigma_{pPb}$} (nb/nucleon)&0.04 $\underline{+}$ 0.07 & 17.99 $\underline{+}$ 10.78 & 54.24 $\underline{+}$ 23.42\\
				\hline
				{\bf $\sigma_{pU}$} (nb/nucleon)&0.04 $\underline{+}$ 0.07 & 17.53 $\underline{+}$ 10.60 & 53.04 $\underline{+}$ 23.20\\
				\hline
			\end{tabular}
		}
		\caption{ Extrapolated values of $\sigma_{abs}^{J/\psi}$ and per nucleon inclusive $J/\psi$ production cross-section near mid-rapidity~(using Eqn.~\ref{eqn4}) in p+A collisions, for several foreseen nuclear targets and collision energies at the upcoming CBM and NA60+ experiments.}    
		\label{table6}    
	\end{table}

	Till now we have analyzed the inclusive $J/\psi$ production cross sections around mid-rapidity, and the extracted $\sigma_{abs}^{J/\psi}$ values are integrated over available phase space. Further insights on CNM effects may be obtained through the investigation of $J/\psi$ kinematical distributions. At 400 GeV, NA50 experiment observed a constant backward shift of the acceptance corrected centre-of-mass rapidity distribution of the $J/\psi$ mesons, $\Delta y\sim0.2$ (corresponding to $\Delta x_{F}\sim0.045$), and a width $\sigma_{y}=0.85$, independent of the target mass~\cite{arnaldi_NPB}. The corresponding measurements by NA60 experiment at the same energy were found to be fitted with an asymmetric Gaussian function imposing a mean $\mu_{y} = -0.2$ and $\sigma_{y} = 0.81 \pm 0.03$~\cite{arnaldi_NPB}. At 158 GeV, data could be described by a gaussian function centred at mid-rapidity, $\mu_{y} = 0.05 \pm 0.05$ and a width $\sigma_{y} = 0.51 \pm 0.02$~\cite{arnaldi_NPB}~\footnote{The reference~\cite{arnaldi_NPB} is a proceeding article and they contain preliminary results.}. The differential cross section data corresponding to these measurements are yet to be made publicly available. At 450 GeV, NA50 experiment published the per nucleon $x_F$ differential $J/\psi$ production cross section over the range $-0.1 < x_{F} < 0.1$, in four equidistant bins~\cite{NA50_450_GeV}\footnote{Unlike the integrated cross section, the reanalyzed differential cross sections, consistent with 400 GeV measurements, were not published.}. No significant dependence of the effective dissociation cross section was observed. The corresponding $\sigma_{abs}^{J/\psi}(x_{F})$ values extracted via both full Glauber analysis and $\rho L$ parametrization, were found to be the same within errors. Instead of differential production cross sections, NA60 published the $x_F$ dependence of $\alpha_{J/\psi}$ subdivided in 5(4) $x_F$ bins at 158(400) GeV. $\alpha_{J/\psi}$ was seen to decrease steadily while going from negative towards positive $x_{F}$, an effect already seen at higher energies. Ref.~\cite{scomparin_NPA} presents a comparison of $\alpha_{J/\psi}$ values as a function of $x_F$ from different fixed target p+A experiments between 158 GeV to 920 GeV. The corresponding values of $\sigma_{abs}^{J/\psi}$\footnote{calculated from the $x_F$ dependence of $\alpha_{J/\psi}$} are presented in Ref.~\cite{arnaldi_NPB}. At a given collision energy, stronger nuclear effects (smaller $\alpha_{J/\psi}$ or larger $\sigma_{abs}^{J/\psi}$) at higher $x_F$ indicates the presence of additional absorption mechanisms at forward rapidities presumably related to energy loss and formation time effects. Near the mid-rapidity, in the region $x_F< 0.25$ the measured suppression pattern is rather flat indicating that the mechanisms causing stronger absorption at forward $x_F$ are small enough at central rapidity. Restricting our analysis within the pure Glauber framework, to the integrated measurements at midrapidity thus seems to be justified. In Ref.~\cite{lourenco_JHEP}, explicit estimations have shown that if the nuclear modification of the parton densities is ignored, the corresponding $\sigma_{abs}^{J/\psi}$ values are independent of $y_{cms}$ around midrapidity. The presence of nuclear modifications of PDFs, however, causes non-trivial dependence of final state absorption on $J/\psi$ rapidity.
	
	The $J/\psi$ transverse momentum ($p_{T}$) distributions were also measured at SPS. A broadening of the $p_T$ distribution as a function of target mass was observed by the NA60 experiment both at 158 and 400 GeV, in line with the results from previous experiment and usually attributed to the initial state multiple scattering of the incoming gluons before they undergo hard scattering to produce a $c\bar{c}$ pair. The target mass dependence of the mean $p_T$ is parametrized as $<p_{T}^{2}>~=~<p_{T}^{2}>_{pp}~+~\rho(A^{1/3}-1)$, where the last term is roughly proportional to the length~(L) of the nuclear matter crossed by the $c\bar{c}$ pair in its way out of the nucleus. Comparison of data from different fixed target experiments have shown $<p_{T}^{2}>_{pp}$ to grow linearly with square of centre of mass energy s. The slope $\rho$, on the other hand, is almost independent of energy apart from a hint of steep fall at low energy, corresponding to NA60 measurement at 158 GeV. Below 158 GeV, slope would be further down, which might lead to a $<p_{T}>$ broadening independent of the target mass~\cite{arnaldi_NPB}.

	The unavailability of appropriate data at SPS energies limits us to extend our analysis to differential studies.

	We now move to heavy-ion collisions at the lower energies. As previously mentioned, in A+A collisions, we only include the expected magnitude of the CNM effect via the prescription discussed above. The $J/\psi$ production cross-section in nucleus-nucleus~(A+B) collisions are estimated using the $\rho L$ parameterizations by changing the L to L$_A$+L$_B$ in Eqn.~\ref{eqn4} to account for the nucleons present in both target and projectile nuclei. Any possible centrality dependence of $\sigma_{abs}^{J/\psi}$ is ignored and the effective $J/\psi$ absorption cross section in nuclear collisions is assumed to be the same as the value estimated in p+A collisions at the same energy, as being done in previous analyses at SPS. The validity of this assumption also demands the same kinematic coverage of the p+A and A+A measurements. However as discussed earlier, once the initial state modification of the parton densities (shadowing/antishadowing corrections) are explicitly taken into account, the $\sigma_{abs}^{J/\psi}$ extracted from Glauber analysis of p+A collisions denotes the final state absorption cross-section of the $J/\psi$ mesons in the nuclear matter of the target nucleus. In p+A collisions, only partons inside the target nucleus are influenced by shadowing effects while in nuclear collisions modifications of both target and projectile partons must be considered. Neglecting shadowing in p+A to A+A extrapolations, introduces a small bias, resulting in an artificial suppression of $J/\psi$ yield, the specific amount of which would depend on the opted parametrization of the nPDF, as shown in Ref.~\cite{arnaldi_prc}, for heavy-ion data at SPS\footnote{In fact in heavy-ion collisions the shadowing effects might be different for various centrality interval, more important for the core than in halo. This might also lead to a centrality dependent local final state absorption.}.
	\begin{table} [h!]
		{
			\begin{tabular}{ |c|c|c|c|}
				\hline
				{\bf $\sqrt{s_{NN}}$}&7.6 GeV&12.3 GeV&17.0 GeV\\ 
				\hline 
				{\bf $\sigma_{abs}^{J/\psi}$ } (mb)&8.51 $\underline{+}$ 1.40 & 7.47 $\underline{+}$ 1.48 & 6.44 $\underline{+}$ 1.60\\
				\hline 
				{\bf $\sigma_{NiNi}$} (nb/nucleon)&0.04 $\underline{+}$ 0.06 & 15.24 $\underline{+}$ 9.64 & 47.03 $\underline{+}$ 21.97\\
				\hline
				{\bf $\sigma_{AuAu}$} (nb/nucleon)&0.02 $\underline{+}$ 0.04 & 10.55 $\underline{+}$ 7.44 & 34.23 $\underline{+}$ 18.69\\
				\hline
				{\bf $\sigma_{PbPb}$} (nb/nucleon)&0.02 $\underline{+}$ 0.04 & 10.34 $\underline{+}$ 7.33 & 33.64 $\underline{+}$ 18.52\\
				\hline
			\end{tabular}
		}
		\caption{ Extrapolated values of $J/\psi$ production cross-section in nucleus-nucleus collisions (A+A) for three different systems at foreseeable CBM and NA60+ energies using $\rho L$ parameterization.}    
		\label{table7b}   
	\end{table}
\begin{figure}[htb!]
	\includegraphics[width=0.49\textwidth]{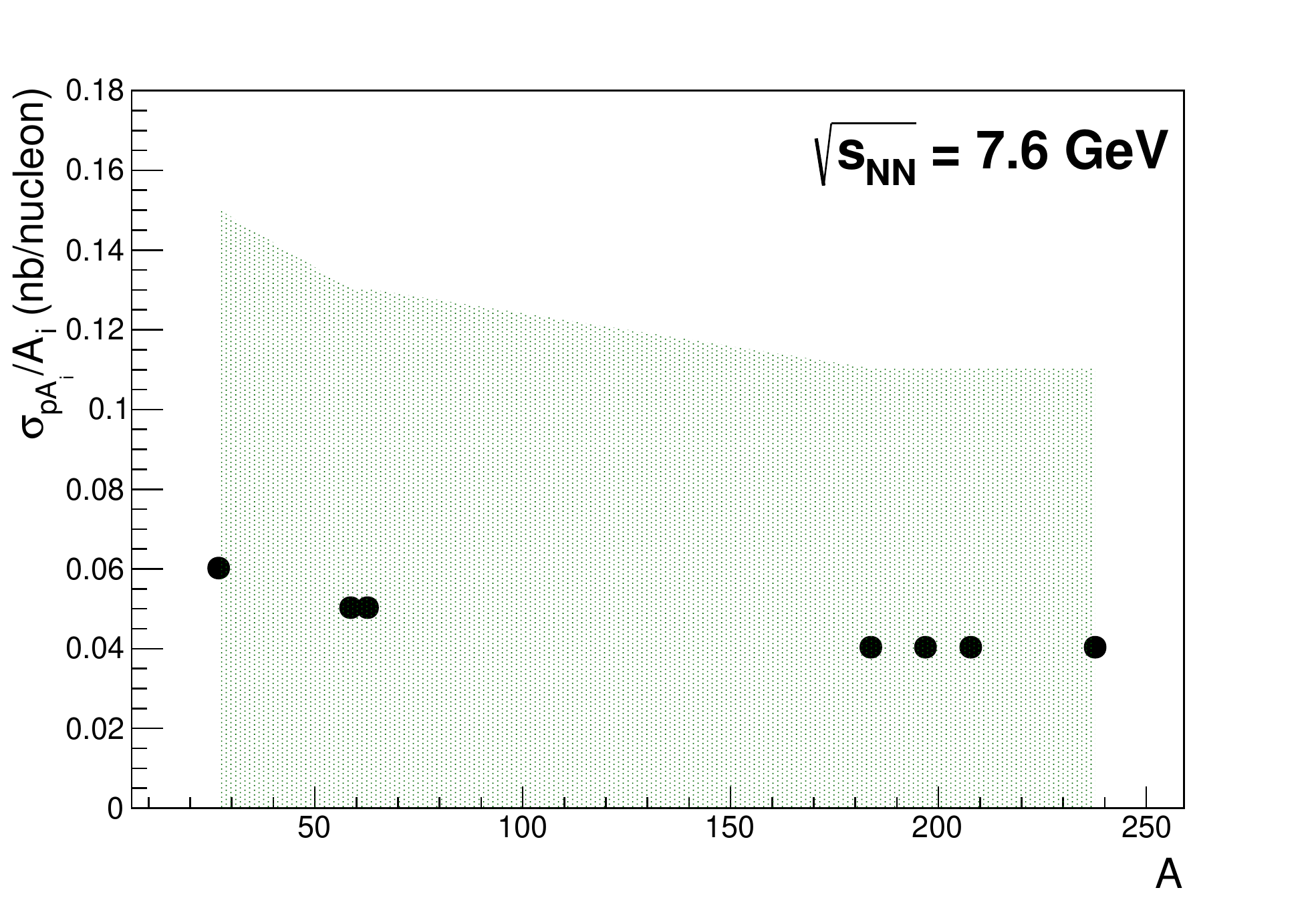}
	\includegraphics[width=0.49\textwidth]{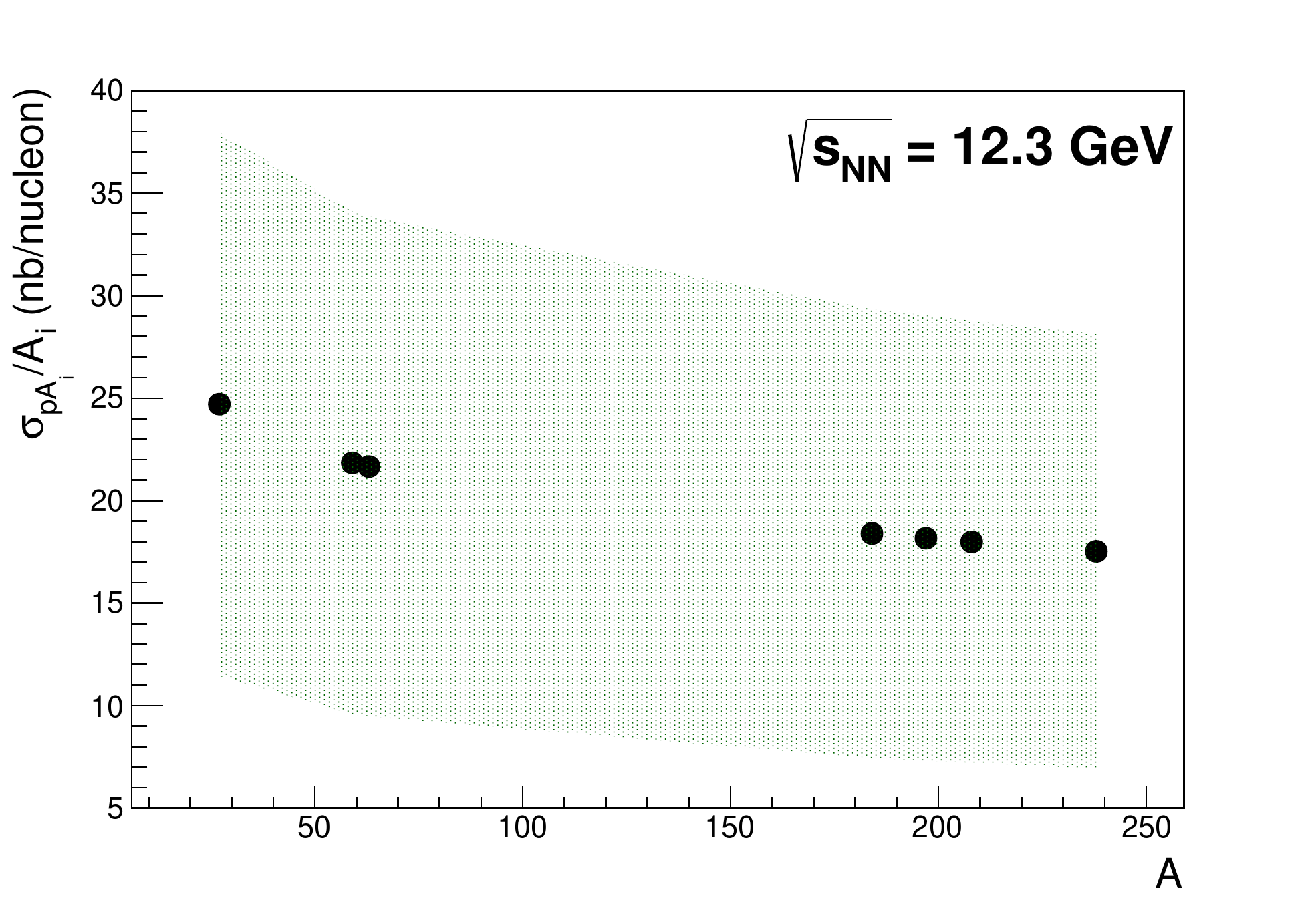}
	\includegraphics[width=0.49\textwidth]{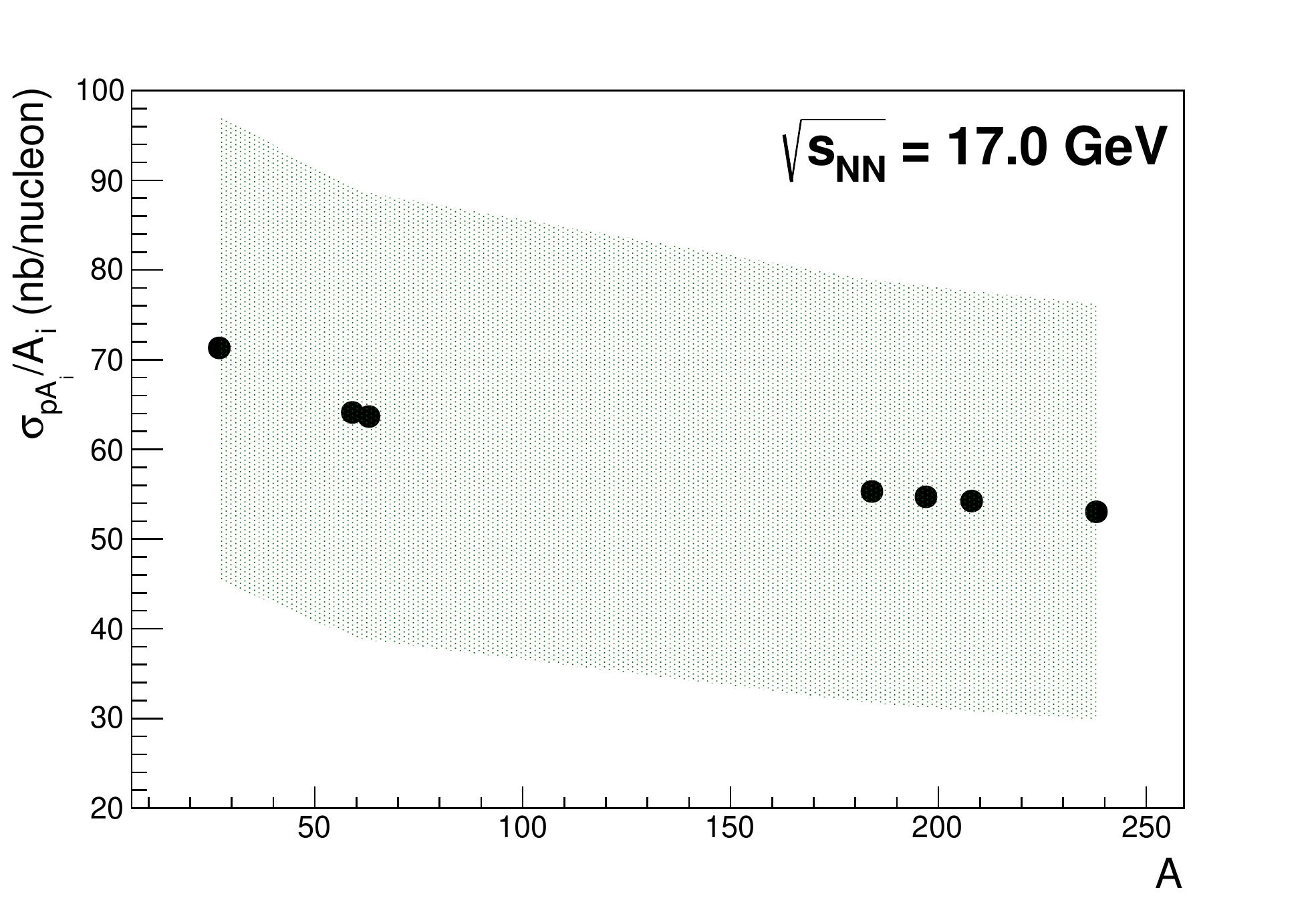}
	\caption{Variation of the  per nucleon $J/\psi$ production cross-section in p+A collisions, as a function of target mass at three different center of mass energies ($\sqrt{s_{NN}}$) relevant for the upcoming CBM and NA60+ experimental programs.}
	\label{cross_section_pA}
\end{figure}
\begin{figure}[htb!]
	\centering
	\includegraphics[width=0.49\textwidth]{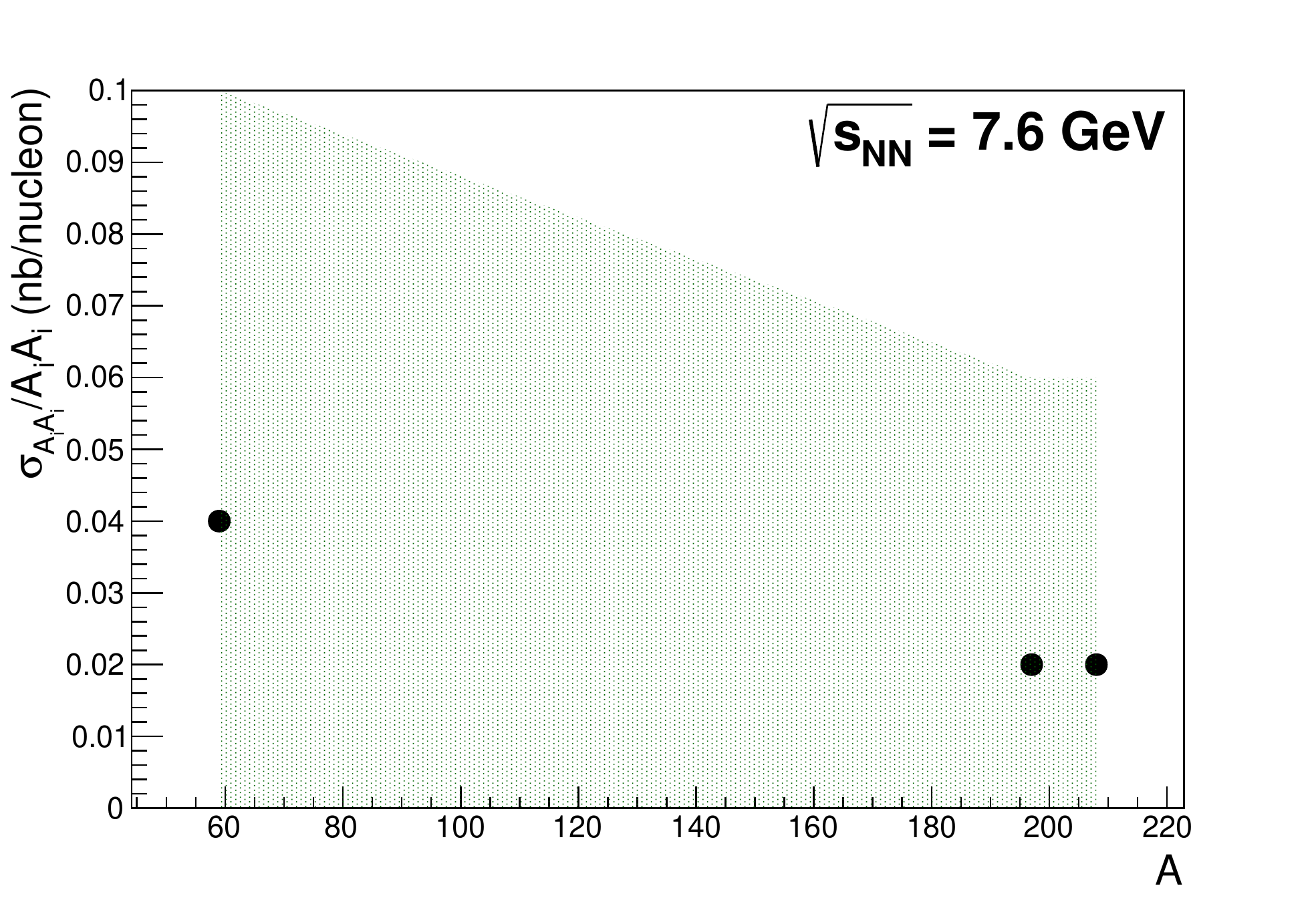}
	\includegraphics[width=0.49\textwidth]{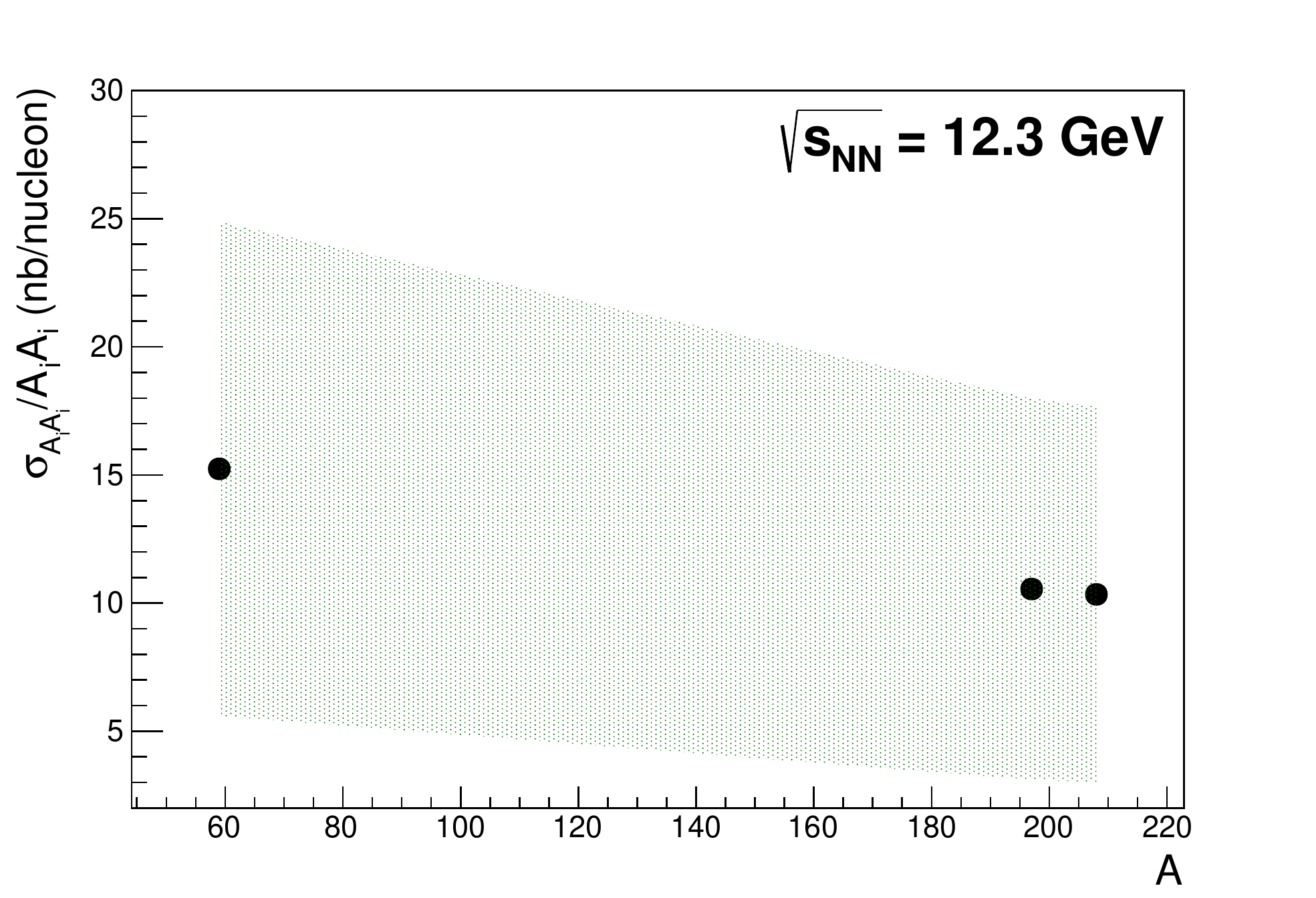}
	\includegraphics[width=0.49\textwidth]{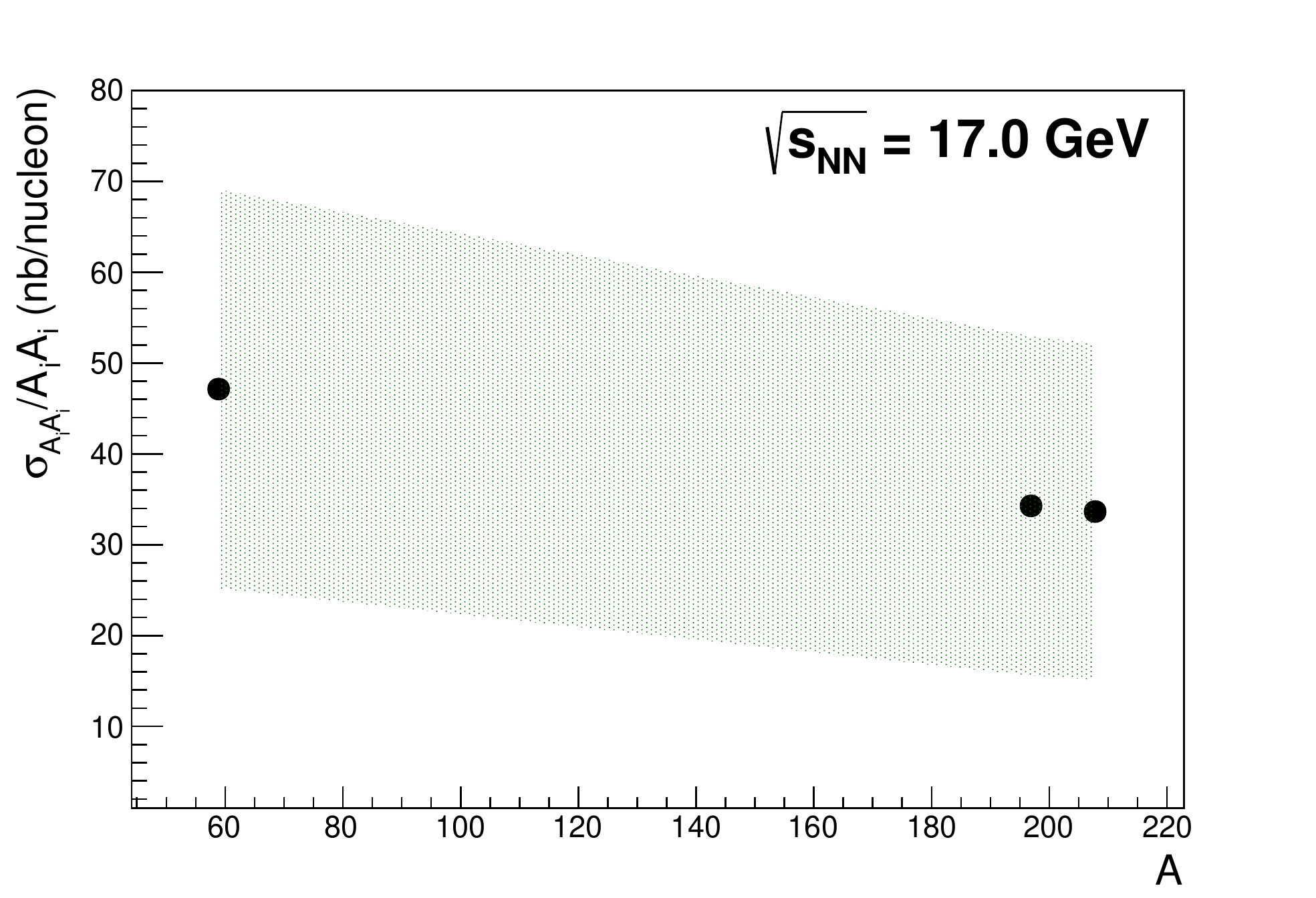}
	\caption{Variation of the $J/\psi$ production cross-section for different nucleus-nucleus collision systems, at three different center of mass energies ($\sqrt{s_{NN}}$) foreseen at the upcoming CBM and NA60+ experiments.}
	\label{cross_section_AA}
\end{figure}

	In Table~\ref{table7b}, the $J/\psi$ production cross-section calculated using $\rho L$ parameterizations are tabulated for nucleus-nucleus collisions. In Fig.~\ref{cross_section_pA} the variation of $J/\psi$ production cross-section in p+A collisions as a function of target mass number is shown at collision energies foreseeable at CBM and NA60+ experiments. The dominant contribution to the uncertainties in the production cross sections originate from the corresponding uncertainty in p+p production, and largest at the lowest investigated energy. The corresponding cross sections for A+A collisions are displayed in Fig.~\ref{cross_section_AA}.
	After calculating the cross-sections, we go on for calculating the yield of $J/\psi$ for different collision systems which will be useful as input for detector performance simulation for the said low energy experiments. 
	
	\section{$J/\psi$ yield in p+A and A+A collisions}

	At an impact parameter b, the $J/\psi$ yield in p+A collisions can be written as 
	\begin{equation}
	\langle N_{\mbox{\scriptsize{\it{pA}}}}^{J/\psi}\rangle (b) \;=\; 
	\sigma_{\mbox{\scriptsize{\it{pA}}}}^{J/\psi}\cdot T_{\mbox{\scriptsize{\it{pA}}}}(b)\;
	\label{eqn:7}
	\end{equation}
	where $\sigma_{pA}^{J/\psi}$ denotes the $J/\psi$ production cross-section in p+A collisions and T$_{pA}(b)$ is the nuclear thickness function at an impact parameter b. 
	The corresponding mean number of binary collisions is given by:
	\begin{equation}
	\langle N_{coll}\rangle (b) \; = \; \sigma_{\mbox{\scriptsize{\it{NN}}}}\cdot T_{\mbox{\scriptsize{\it{pA}}}}(b)
	\label{eqn:8}
	\end{equation}
	where $\sigma_{NN}$ is the nucleon-nucleon inelastic interaction cross-section. 
	The collision energy $\sqrt{s}$ dependence of $\sigma_{NN}$ is parameterized using a polynomial function of the following form~\cite{david_d_enteria}, 
	\begin{equation}
	a+b\times ln^{n}(s)
	\label{eqn_6}
	\end{equation}
	where a~=~28.84~$\underline{+}~0.52$, b~=~0.05~$\underline{+}~0.02$ and c~=~2.37~$\underline{+}~0.12$
	

	From Eqn.~\ref{eqn:8} it appears that the nuclear overlap function, 
	$T_{\mbox{\scriptsize{\it{pA}}}}(b)~=~N_{coll}(b)/\sigma_{\mbox{\scriptsize{\it{NN}}}}$, can be considered as the luminosity (reaction rate per unit of cross-section) per collision 
	at an impact parameter b. For practical purposes, a more useful estimate is the $J/\psi$ yield in p+A collisions for a given centrality interval, $(N_{pA}^{J/\psi})_{C_1-C_2}$. 
	In order to do so, let us define two parameters~\cite{r_vogt}
	
	\begin{itemize}
		\item The fraction of the total cross-section for $J/\psi$ production over an impact parameter range
		$b_{1}<b<b_{2}$  ($d^2b = 2\pi bdb$):
		\begin{equation}
		f_{J/\psi}(b_1<b<b_2)\;=\;\frac{2\pi}{A}\int_{b_1}^{b_2} bdb\; T_{\mbox{\scriptsize{\it{pA}}}}(b).
		\label{eq:f_AB}
		\end{equation}
		\item The fraction of the geometric cross-section within the same range 
		\begin{equation}
		f_{geo}(b_1<b<b_2)\;=\;\left [2\pi\; \int_{b_1}^{b_2} bdb \left(1-e^{-\sigma_{\mbox{\scriptsize{\it{NN}}}}
			T_{\mbox{\scriptsize{\it{pA}}}}(b)}\right)\right]/\sigma_{\mbox{\scriptsize{\it{pA}}}}^{geo}\; ,
		\label{eq:f_geo}
		\end{equation}
		
	\end{itemize}

	Thus the nuclear thickness function for any given centrality class $C_1-C_2$ can be expressed as,
	
	\begin{equation}
	\langle T_{\mbox{\scriptsize{\it{pA}}}}\rangle_{C_1-C_2} \;\equiv\; \frac{\int^{b_2}_{b_1} d^2b \; T_{\mbox{\scriptsize{\it{pA}}}}}
	{\int^{b_2}_{b_1} d^2b} = \frac{A}{\sigma_{\mbox{\scriptsize{\it{pA}}}}^{geo}} \cdot \frac{f_{J/\psi}}{f_{geo}}\;
	\label{eqn:11}
	\end{equation}
	where $\sigma_{pA}^{geo}=\int d^{2}b~[1-e^{-\sigma_{NN}T_{pA}(b)}]$. It is clear from the Eqn.~\ref{eqn:11} that for minimum bias collision~(i.e. integrated over the all possible values of impact parameters), $\langle T_{pA} \rangle _{MB}~=~\frac{A}{\sigma_{pA}^{geo}}$. Finally, the $J/\psi$ yield for any centrality~(C$_1$-C$_2$) class can be expressed as,

	\begin{equation}
	\langle N_{\mbox{\scriptsize{\it{pA}}}}^{J/\psi}\rangle_{C_{1}-C_{2}} \;=\;  \sigma_{\mbox{\scriptsize{\it{pA}}}}^{J/\psi}\cdot \langle T_{\mbox{\scriptsize{\it{pA}}}}\rangle_{C_{1}-C_{2}}\;,
	\label{eqn:12}
	\end{equation} 
	Though the centrality dependence of the nuclear thickness function is calculated using the ratio of f$_{J/\psi}$/f$_{geo}$ but the centrality dependence of the nuclear absorption cross-section is not taken into account in the course of this exercise.
	
\begin{figure}[htb!]
		\centering
		\includegraphics[width=0.65\textwidth, height=0.30\textheight]{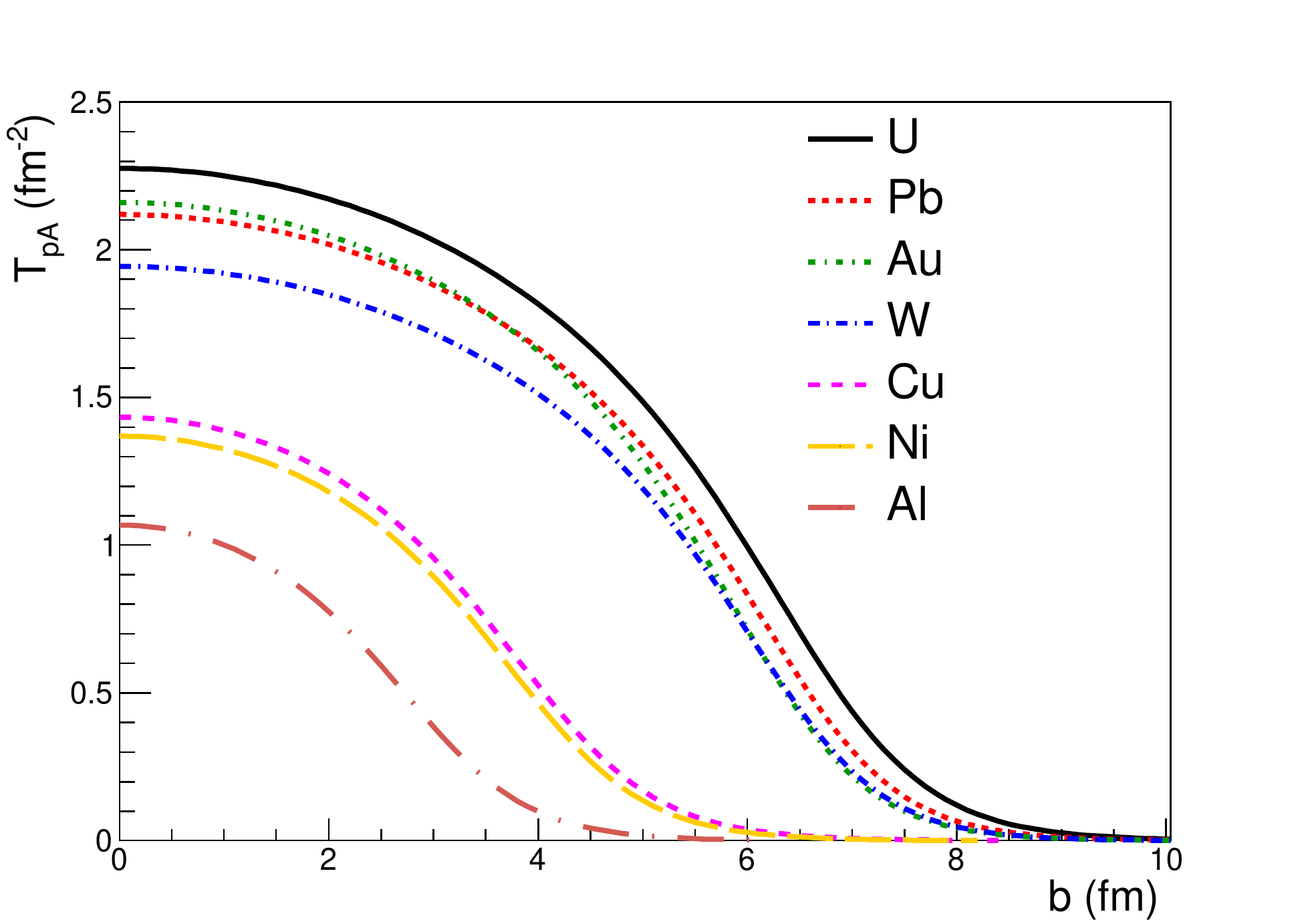}
		\caption{Impact parameter~(b) dependence of nuclear thickness function  for different nuclei.}
		\label{thickness_function}
	\end{figure}

	The distribution of the nuclear thickness function for the different nuclei is shown in Fig.~\ref{thickness_function}. The nuclear density is parameterized by a Woods-Saxon distribution and the respective nuclear shape parameters are taken from Ref.~\cite{Vogt,nuclear_data_book,nuclear_data_book_1}.
	\begin{figure}[htb!]
		\centering
		\includegraphics[width=0.65\textwidth, height=0.30\textheight]{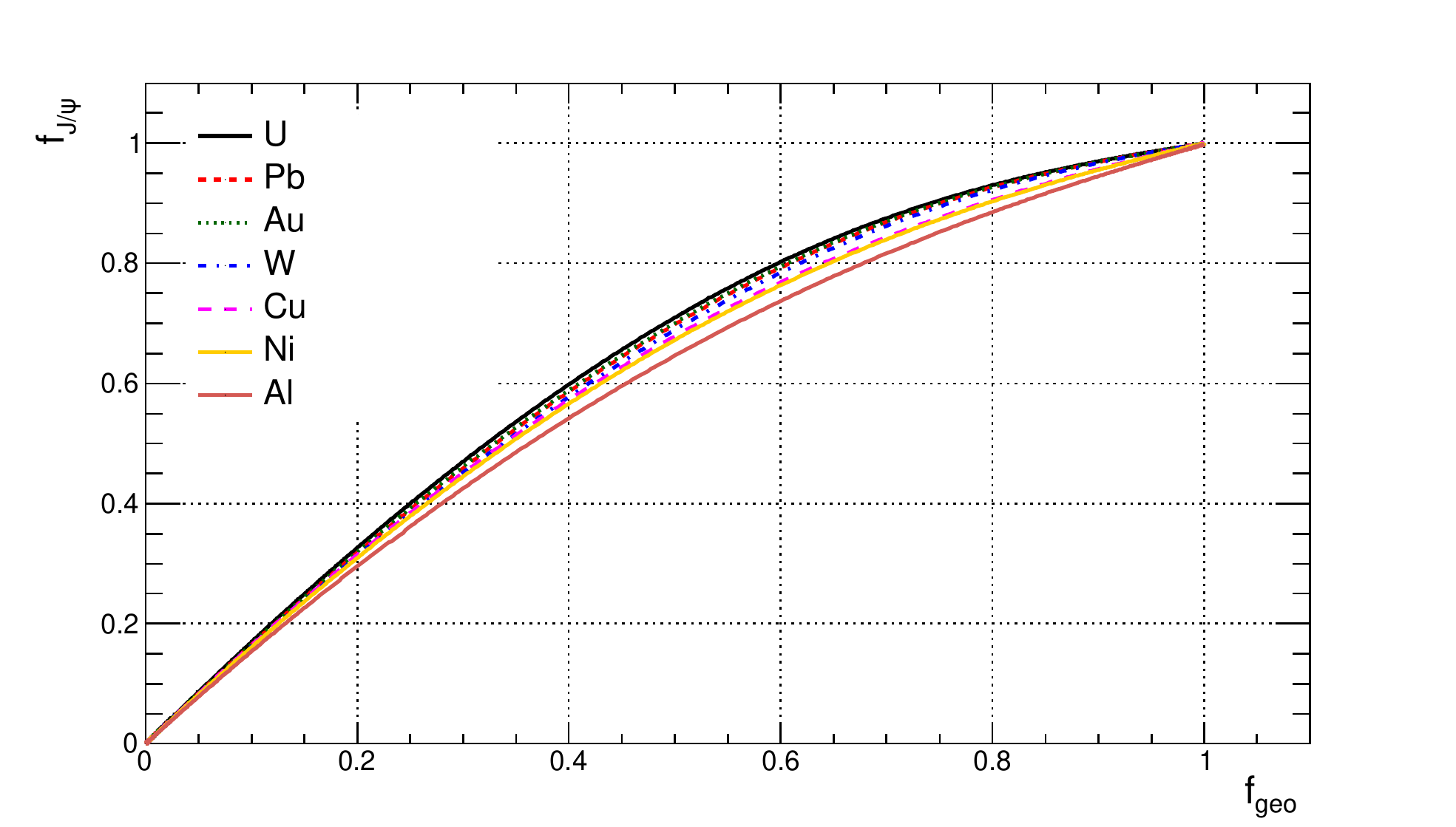}
		\caption{Variation of f$_{J/\psi}$ as a function of f$_{geo}$ for different proton-nucleus system at $\sqrt{s}~=7.62~GeV$~with $\sigma_{NN} \approx 30~mb$}
		\label{f_geo_f_hard_pA}
	\end{figure}
	\begin{table}[htb!]
		\centering
		\resizebox{\columnwidth}{!}
		{
			\begin{tabular}{|c|c|c|c|c|c|c|c|c|}
				\hline
				{ Centrality} & {f$_{geo}$} & {f$_{J/\psi}$} &  {f$_{J/\psi}$} & {f$_{J/\psi}$}& {f$_{J/\psi}$}&{f$_{J/\psi}$}& {f$_{J/\psi}$} & {f$_{J/\psi}$}  \\
				{(C$_{1}$ - C$_{2}$)} &  & { (p+Al)} & { (p+Ni)} & {(p+Cu)} & { (p+W)}& {(p+Au)} & { (p+Pb)}& { (p+U)}\\
				\hline
				0-20\% &0.20& 0.30& 0.32 &0.32 &   0.32& 0.33& 0.32& 0.33\\ 
				\hline
				20-40\%& 0.20&0.25& 0.26 &0.26  & 0.27& 0.27& 0.27 & 0.28 \\ 
				\hline
				40-60\%& 0.20&0.19& 0.20 &0.20& 0.21& 0.21& 0.21 & 0.20 \\ 
				\hline
				60-80\%& 0.20 &0.14& 0.14&0.13&  0.13& 0.13& 0.13 & 0.12 \\ 
				\hline
				80-100\%& 0.20 &0.11& 0.09&0.09& 0.07& 0.07 & 0.07& 0.06 \\ 
				\hline
			\end{tabular}
		}
		\caption{The fraction of total cross-section for hard processes and the fraction of geometric cross-section for given centrality classes in p+A collisions for different collision systems.}		
		\label{table10}
	\end{table}	
	
	\begin{table}[htb!]
		\centering
		{
			\begin{tabular}{|c|c|c|c|}
				\hline
				{$\sqrt{s_{NN}}$ (GeV)}&{ System (p+A)} & { $\sigma^{geo}_{pA}$ (fm$^{2}$)} & { $<T_{pA}>_{MB}$ (fm$^{-2}$)} \\
				\hline
				&p+Al& 37.90~$\underline{+}$ 4.99& 0.71~$\underline{+}$ 0.10\\ \cline{2-4}
				&p+Ni&68.89~$\underline{+}$ 7.32& 0.86~$\underline{+}$ 0.09\\ \cline{2-4}				
				&p+Cu& 73.58~$\underline{+}$ 7.70 & 0.86~$\underline{+}$ 0.09  \\ \cline{2-4}
				7.6&p+W& 157.18~$\underline{+}$ 10.72 & 1.17~$\underline{+}$ 0.08 \\ \cline{2-4}
				&p+Au& 155.75~$\underline{+}$10.02 & 1.26~$\underline{+}$ 0.08 \\ \cline{2-4}
				&p+Pb& 166.53~$\underline{+}$10.77 & 1.25~$\underline{+}$ 0.08 \\ \cline{2-4}
				&p+U& 182.94~$\underline{+}$11.77  & 1.30~$\underline{+}$ 0.08 \\ \cline{2-4}
				\hline
				&p+Al& 38.44~$\underline{+}$ 6.95& 0.70~$\underline{+}$ 0.13\\ \cline{2-4}
				&p+Ni&69.66~$\underline{+}$ 10.23& 0.85~$\underline{+}$ 0.13\\ \cline{2-4}				
				&p+Cu& 74.39~$\underline{+}$ 10.76 & 0.85~$\underline{+}$ 0.12  \\ \cline{2-4}
				12.3&p+W& 158.29~$\underline{+}$ 15.10 & 1.16~$\underline{+}$ 0.11 \\ \cline{2-4}
				&p+Au& 156.79~$\underline{+}$ 14.12 & 1.26~$\underline{+}$ 0.11 \\ \cline{2-4}
				&p+Pb& 167.65~$\underline{+}$ 15.20 & 1.24~$\underline{+}$ 0.11 \\ \cline{2-4}
				&p+U& 184.12~$\underline{+}$ 16.59 & 1.29~$\underline{+}$ 0.12 \\ \cline{2-4}
				\hline
				&p+Al& 38.90~$\underline{+}$ 8.95& 0.69~$\underline{+}$ 0.17 \\ \cline{2-4}
				&p+Ni& 70.33~$\underline{+}$ 13.28& 0.84~$\underline{+}$ 0.16 \\ \cline{2-4}				
				&p+Cu& 75.09~$\underline{+}$ 13.98 &0.84~$\underline{+}$ 0.16   \\ \cline{2-4}
				17.0&p+W& 159.24~$\underline{+}$ 19.88 & 1.16~$\underline{+}$ 0.15  \\ \cline{2-4}
				&p+Au& 157.68~$\underline{+}$ 18.61 & 1.25~$\underline{+}$ 0.15  \\ \cline{2-4}
				&p+Pb& 168.61~$\underline{+}$ 20.01 & 1.23~$\underline{+}$ 0.15  \\ \cline{2-4}
				&p+U& 185.22~$\underline{+}$ 21.83 & 1.28~$\underline{+}$ 0.15	 \\ \cline{2-4}
				\hline
			\end{tabular}
		}
		\caption{The geometric cross-section, average nuclear thickness function applicable for a any hard process in p+A collisions for a variety of target nuclei at different collision energies.}		
		\label{table11}
	\end{table}
	The $J/\psi$~(hard) production is more enhanced for central collisions as compared to the total reaction cross-section. The growth of the geometric cross-section as a function of collision centrality is slower than that of the hard component. The variation of f$_{J/\psi}$ as a function of f$_{geo}$ for different systems are shown in Fig.~\ref{f_geo_f_hard_pA}. In Table~\ref{table10}, the values of the fractions (f$_{J/\psi}$ and f$_{geo}$) at different centralities for different p+A collision systems are given. The values of the nuclear thickness function for minimum bias collision systems and the geometric cross-sections for the p+A collision systems are tabulated in Table~\ref{table11} for different collision energies.
	
	A similar exercise can be performed for nucleus-nucleus~(A+B) collisions systems by replacing pA with AB. Thus the nuclear overlap function for any given centrality class $C_1-C_2$ can be expressed as,
	
	\begin{equation}
	\langle T_{\mbox{\scriptsize{\it{AB}}}}\rangle_{C_1-C_2} \;\equiv\; \frac{\int^{b_2}_{b_1} d^2b \; T_{\mbox{\scriptsize{\it{AB}}}}}
	{\int^{b_2}_{b_1} d^2b} = \frac{AB}{\sigma_{\mbox{\scriptsize{\it{AB}}}}^{geo}} \cdot \frac{f_{J/\psi}}{f_{geo}}\;
	\label{eqn:13}
	\end{equation}
	where $\sigma_{AB}^{geo}=\int d^{2}b~[1-e^{-\sigma_{NN}T_{AB}(b)}]$. Thus the $J/\psi$ yield for any centrality~(C$_1$-C$_2$) class can be expressed as,
	
	\begin{figure*}[htb!]
		\centering
		\includegraphics[width=0.8\textwidth]{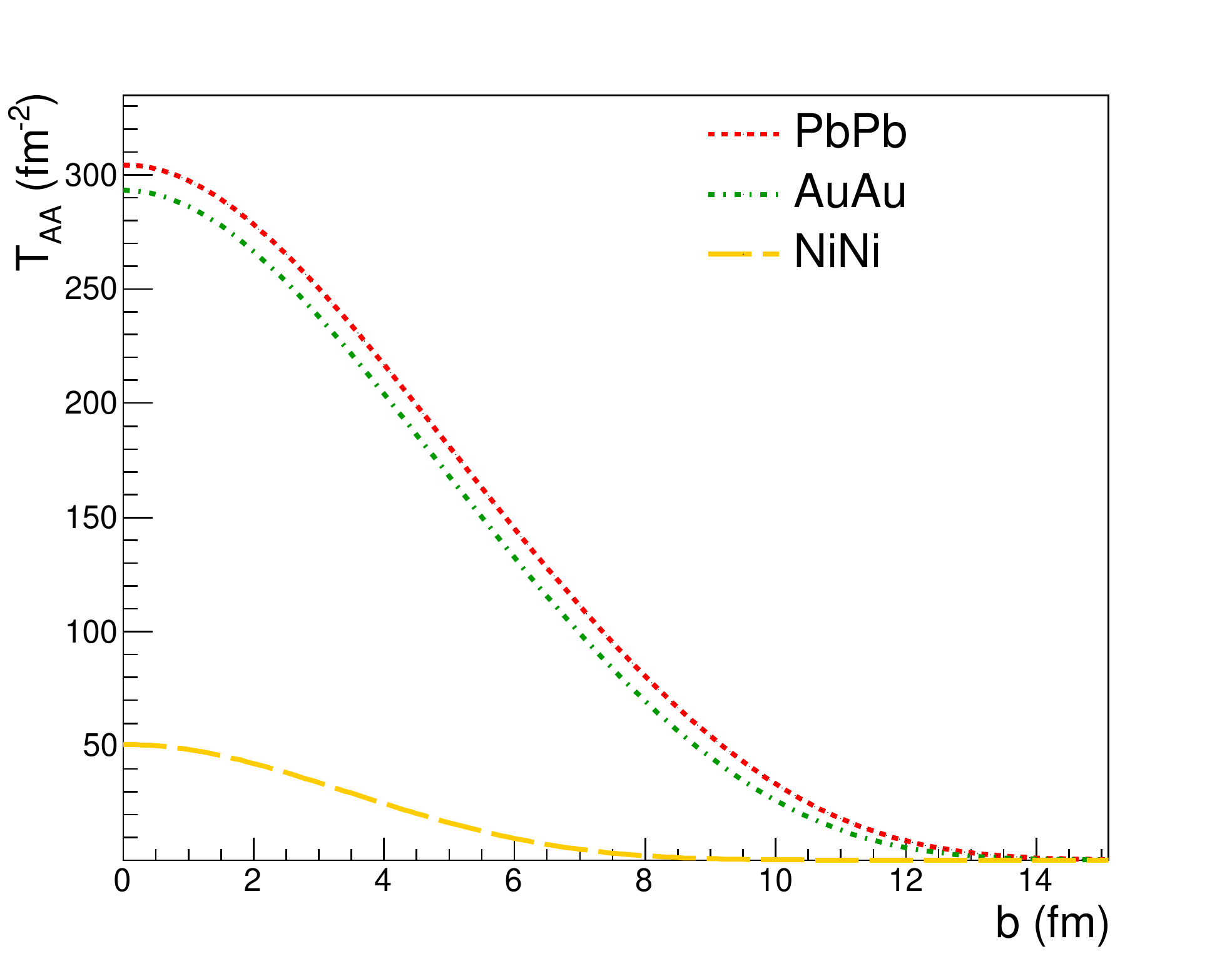}
		\caption{Variation of nuclear overlap function with impact parameter~(b) for different nuclei.}
		\label{fig9}
	\end{figure*}
	
	\begin{equation}
	\langle N_{\mbox{\scriptsize{\it{AB}}}}^{J/\psi}\rangle_{C_{1}-C_{2}} \;=\;  \sigma_{\mbox{\scriptsize{\it{AB}}}}^{J/\psi}\cdot \langle T_{\mbox{\scriptsize{\it{AB}}}}\rangle_{C_{1}-C_{2}}\;,
	\label{eqn:14}
	\end{equation} 
	The distribution of the nuclear overlap function with impact parameter b, for the different colliding nuclei is shown in Fig.~\ref{fig9}. The uncertainties in the overlap function might arise due to the possible uncertainties in the nuclear density profile.

\begin{figure*}[htb!]
		\centering
		\includegraphics[width=0.8\textwidth]{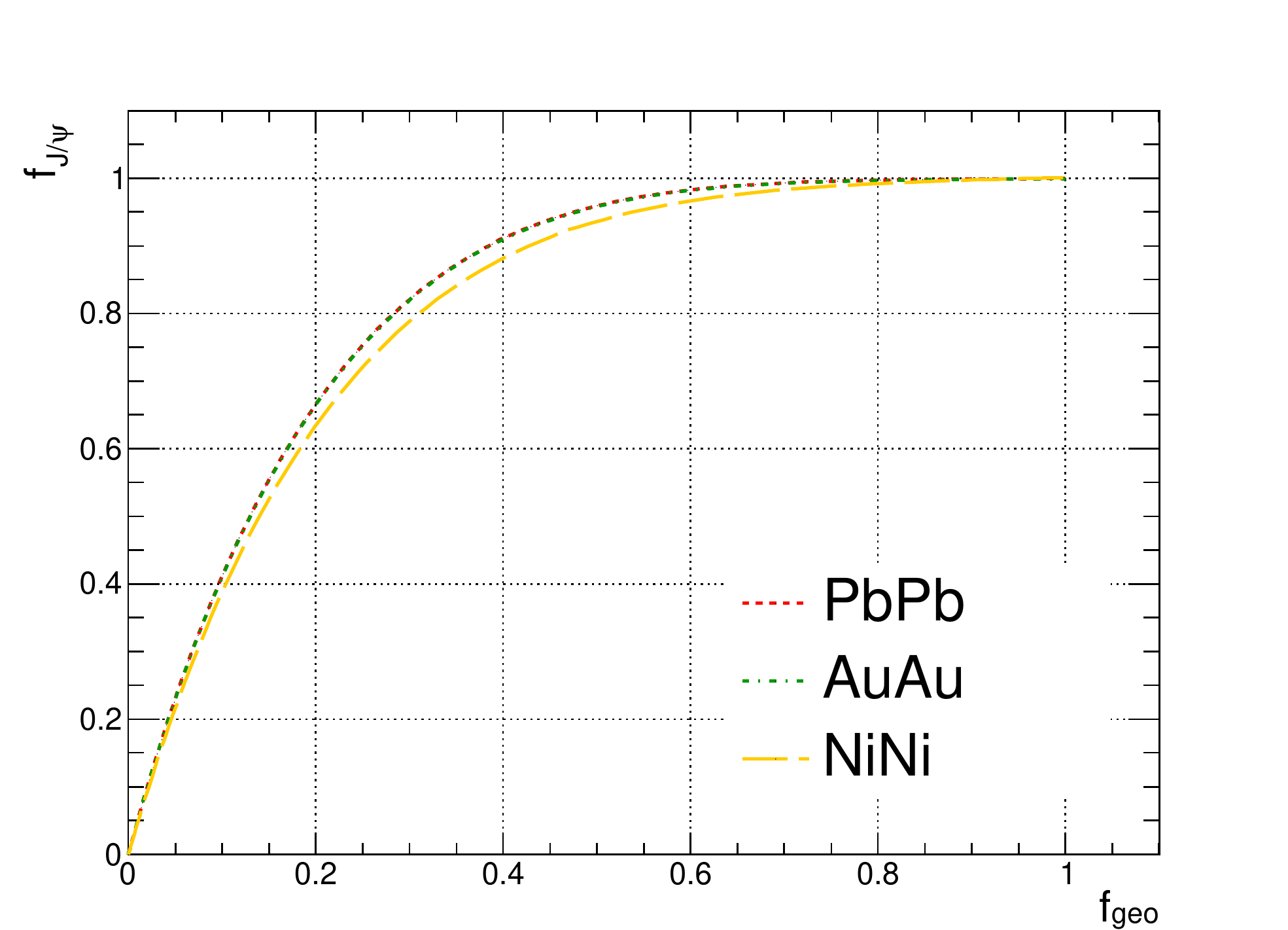}
		\caption{Variation of f$_{J/\psi}$ as a function of f$_{geo}$ for different proton-nucleus system at $\sqrt{s}~=7.62~GeV$~with $\sigma_{NN} \approx 30~mb$}
		\label{fig10}
	\end{figure*}

	The variation of f$_{J/\psi}$ as a function of f$_{geo}$ for different A+A collision systems is shown in Fig.~\ref{fig10}. In Table~\ref{table_AA_fraction_geo_hard_AA}, the values of the fractions (f$_{J/\psi}$ and f$_{geo}$) at different centralities for different A+A collision systems are given. The values of the nuclear overlap function for minimum bias collision systems and the geometric cross-sections for the A+A collision systems are tabulated in Table~\ref{table13} for different collision energies.

	\begin{table}[htb!]
		\centering
		{
			\begin{tabular}{|c|c|c|c|c|}
				\hline
				{ Centrality} & {f$_{geo}$} & {f$_{J/\psi}$} &{f$_{J/\psi}$}& {f$_{J/\psi}$}  \\
				{(C$_{1}$ - C$_{2}$)} &  & {(Ni+Ni)}&  {(Au+Au)}& {(Pb+Pb)}\\
				\hline
				0-20\%&0.20& 0.63  & 0.66& 0.67\\
				\hline
				20-40\%& 0.20 & 0.25 & 0.24& 0.24\\
				\hline
				40-60\%& 0.20 & 0.09 & 0.07& 0.07\\
				\hline
				60-80\%& 0.20 & 0.03 & 0.02& 0.01\\
				\hline
				80-100\%& 0.20 & 0.01 & 0.002& 0.002\\
				\hline
			\end{tabular}
		}
		\caption{The fraction of total cross-section for hard processes and the fraction of geometric cross-section for given centrality classes in A+A collision for different nuclei.}		
		
		\label{table_AA_fraction_geo_hard_AA}
	\end{table}
	\begin{table*}[htb!]
		\centering
		
		{
			\begin{tabular}{|c|c|c|c|}
				\hline
				{$\sqrt{s_{NN}}$ (GeV)}&{ System (A+A)} & { $\sigma^{geo}_{AB}$ (fm$^{2}$)} & { $<T_{AB}>_{MB}$ (fm$^{-2}$)} \\
				\hline
				&Ni+Ni& 316.32~$\underline{+}$ 13.24& 11.01~$\underline{+}$ 0.46\\ \cline{2-4}				
				7.6&Au+Au& 685.43~$\underline{+}$18.37 & 56.62~$\underline{+}$ 1.52 \\ \cline{2-4}
				&Pb+Pb& 738.18~$\underline{+}$19.88 & 58.61~$\underline{+}$ 1.58 \\ \cline{2-4}
				\hline
				&Ni+Ni&317.73~$\underline{+}$ 18.52& 10.96~$\underline{+}$ 0.64\\ \cline{2-4}				
				12.3&Au+Au& 687.40~$\underline{+}$ 25.70 & 56.46~$\underline{+}$ 2.11 \\ \cline{2-4}
				&Pb+Pb& 740.30~$\underline{+}$ 27.82 & 58.44~$\underline{+}$ 2.20 \\ \cline{2-4}
				\hline
				&Ni+Ni& 318.95~$\underline{+}$ 24.02& 10.91~$\underline{+}$ 0.83 \\ \cline{2-4}				
				17.0&Au+Au& 689.09~$\underline{+}$ 33.33 & 56.32~$\underline{+}$ 2.73  \\ \cline{2-4}
				&Pb+Pb& 742.14~$\underline{+}$ 36.07 & 58.30~$\underline{+}$ 2.84  \\ \cline{2-4}
				\hline
			\end{tabular}
		}
		\caption{The geometric cross-section and the average nuclear overlap function in  minimum bias A+A collisions, for any given hard process at different collision energies and for different colliding systems.}		
		\label{table13}
	\end{table*}
	
	In Table~\ref{table14} and Table~\ref{table15}, the calculated $J/\psi$ yield is summarised for different collision systems at different collision energies and centralities in p+A and A+A collision systems respectively. Though the centrality differential predictions for p+A collisions are quoted here using a data driven approach but these measurements are not simple due to limited resolution of centrality in p+A collisions as well as the correlation between observable and estimator. As earlier stated, the large errors associated with the predicted production yields arise from the large uncertainties in the corresponding p+p production cross sections. A precise estimation of the $J/\psi$ production cross section in p+p collisions is thus highly welcome and would certainly reduce the present uncertainties in the predicted results for p+A and A+A collisions.

	\vspace*{-1.0cm}
	\begin{table*}[htb!]
		\centering
		\resizebox{1.2\columnwidth}{!}
		{
			\hspace*{-3.0cm}
			\begin{tabular}{|c|c|c|c|c|c|c|c|c|}
				\hline
				{ $\sqrt{s_{NN}}$} & { centrality} & {$<N_{pAl}^{J/\psi}>$} & {$<N_{pNi}^{J/\psi}>$}  & {$<N_{pCu}^{J/\psi}>$} & {$<N_{pW}^{J/\psi}>$} & {$<N_{pAu}^{J/\psi}>$}& {$<N_{pPb}^{J/\psi}>$} & {$<N_{pU}^{J/\psi}>$} \\
				(GeV)&(\%)&&&&&&&\\
				\hline
				& 0-20& 6.1$\times 10^{-9}$ & 6.5$\times 10^{-9}$ &6.6$\times 10^{-9}$ &7.3$\times 10^{-9}$ & 8.2$\times 10^{-9}$ & 7.8$\times 10^{-9}$& 8.4$\times 10^{-9}$  \\
				&&$\underline{+}$9.6$\times 10^{-9}$&$\underline{+}$1.1$\times 10^{-8}$&$\underline{+}$1.1$\times 10^{-8}$&$\underline{+}$1.3$\times 10^{-8}$& $\underline{+}$1.5$\times 10^{-8}$&$\underline{+}$1.4$\times 10^{-8}$&$\underline{+}$1.5$\times 10^{-8}$\\
				\cline{2-9}
				&20-40& 5.1$\times 10^{-9}$ &5.3$\times 10^{-9}$ &5.4$\times 10^{-9}$&6.2$\times 10^{-9}$&6.7$\times 10^{-9}$&6.6$\times 10^{-9}$&7.1$\times 10^{-9}$ \\	
				&&$\underline{+}$8.0$\times 10^{-9}$&$\underline{+}$8.9$\times 10^{-9}$&$\underline{+}$8.9$\times 10^{-9}$&$\underline{+}$1.1$\times 10^{-8}$& $\underline{+}$1.2$\times 10^{-8}$&$\underline{+}$1.2$\times 10^{-8}$&$\underline{+}$1.3$\times 10^{-8}$\\
				\cline{2-9}			
				7.6& 40-60 & 3.9$\times 10^{-9}$ & 4.1$\times 10^{-9}$& 4.1$\times 10^{-9}$& 4.8$\times 10^{-9}$& 5.2$\times 10^{-9}$& 5.1$\times 10^{-9}$& 5.1$\times 10^{-9}$\\
				&&$\underline{+}$6.1$\times 10^{-9}$&$\underline{+}$6.9$\times 10^{-9}$&$\underline{+}$6.9$\times 10^{-9}$&$\underline{+}$8.6$\times 10^{-9}$& $\underline{+}$9.3$\times 10^{-9}$&$\underline{+}$9.2$\times 10^{-9}$&$\underline{+}$9.1$\times 10^{-9}$\\
				\cline{2-9}
				& 60-80 & 2.9$\times 10^{-9}$ & 2.8$\times 10^{-9}$ & 2.7$\times 10^{-9}$& 3.0$\times 10^{-9}$& 3.2$\times 10^{-9}$& 3.2$\times 10^{-9}$& 3.1$\times 10^{-9}$\\
				&&$\underline{+}$4.5$\times 10^{-9}$&$\underline{+}$4.8$\times 10^{-9}$&$\underline{+}$4.5$\times 10^{-9}$&$\underline{+}$5.3$\times 10^{-9}$& $\underline{+}$5.8$\times 10^{-9}$&$\underline{+}$5.7$\times 10^{-9}$&$\underline{+}$5.5$\times 10^{-9}$\\
				\cline{2-9}
				& 80-100 & 2.2$\times 10^{-9}$&1.8$\times 10^{-9}$ & 1.9$\times 10^{-9}$& 1.6$\times 10^{-9}$& 1.7$\times 10^{-9}$& 1.7$\times 10^{-9}$& 1.5$\times 10^{-9}$\\ 
				&&$\underline{+}$3.6$\times 10^{-9}$&$\underline{+}$3.1$\times 10^{-9}$&$\underline{+}$3.1$\times 10^{-9}$&$\underline{+}$2.9$\times 10^{-9}$& $\underline{+}$3.1$\times 10^{-9}$&$\underline{+}$3.1$\times 10^{-9}$&$\underline{+}$2.7$\times 10^{-9}$\\	
				\cline{2-9}
				& minimum & 4.1$\times 10^{-9}$ & 4.1$\times 10^{-9}$&4.1$\times 10^{-9}$&4.6$\times 10^{-9}$&5.0$\times 10^{-9}$&4.9$\times 10^{-9}$& 5.1$\times 10^{-9}$\\ 
				&bias&$\underline{+}$6.4$\times 10^{-9}$&$\underline{+}$6.9$\times 10^{-9}$&$\underline{+}$6.9$\times 10^{-9}$&$\underline{+}$8.2$\times 10^{-9}$& $\underline{+}$8.9$\times 10^{-9}$&$\underline{+}$8.8$\times 10^{-9}$&$\underline{+}$9.1$\times 10^{-9}$\\
				\hline			
				
				& 0-20& 2.4$\times 10^{-6}$& 2.7$\times 10^{-6}$& 2.7$\times 10^{-6}$& 3.3$\times 10^{-6}$& 3.6$\times 10^{-6}$& 3.4$\times 10^{-6}$& 3.6$\times 10^{-6}$\\
				&&$\underline{+}$1.3$\times 10^{-6}$&$\underline{+}$1.6$\times 10^{-6}$&$\underline{+}$1.6$\times 10^{-6}$&$\underline{+}$2.0$\times 10^{-6}$& $\underline{+}$2.2$\times 10^{-6}$&$\underline{+}$2.1$\times 10^{-6}$&$\underline{+}$2.2$\times 10^{-6}$\\
				\cline{2-9}
				&20-40& 2.0$\times 10^{-6}$& 2.2$\times 10^{-6}$& 2.2$\times 10^{-6}$& 2.8$\times 10^{-6}$& 3.0$\times 10^{-6}$& 2.9$\times 10^{-6}$& 3.0$\times 10^{-6}$ \\	
				&&$\underline{+}$1.1$\times 10^{-6}$&$\underline{+}$1.3$\times 10^{-6}$&$\underline{+}$1.3$\times 10^{-6}$&$\underline{+}$1.6$\times 10^{-6}$& $\underline{+}$1.8$\times 10^{-6}$&$\underline{+}$1.7$\times 10^{-6}$&$\underline{+}$1.8$\times 10^{-6}$\\
				\cline{2-9}			
				12.3& 40-60 & 1.5$\times 10^{-6}$ & 1.7$\times 10^{-6}$ & 1.7$\times 10^{-6}$& 2.2$\times 10^{-6}$& 2.3$\times 10^{-6}$& 2.2$\times 10^{-6}$& 2.2$\times 10^{-6}$\\
				&&$\underline{+}$8.2$\times 10^{-7}$&$\underline{+}$9.8$\times 10^{-7}$ &$\underline{+}$9.7$\times 10^{-7}$ &$\underline{+}$1.3$\times 10^{-6}$& $\underline{+}$1.4$\times 10^{-6}$&$\underline{+}$1.4$\times 10^{-6}$&$\underline{+}$1.3$\times 10^{-6}$\\
				\cline{2-9}
				& 60-80 & 1.1$\times 10^{-6}$ & 1.2$\times 10^{-6}$& 1.1$\times 10^{-6}$& 1.3$\times 10^{-6}$& 1.4$\times 10^{-6}$& 1.4$\times 10^{-6}$& 1.3$\times 10^{-6}$ \\
				&&$\underline{+}$6.1$\times 10^{-7}$&$\underline{+}$6.8$\times 10^{-7}$&$\underline{+}$6.3$\times 10^{-7}$&$\underline{+}$7.9$\times 10^{-7}$& $\underline{+}$8.5$\times 10^{-7}$&$\underline{+}$8.4$\times 10^{-7}$&$\underline{+}$7.9$\times 10^{-7}$\\
				\cline{2-9}
				& 80-100 & 8.7$\times 10^{-7}$ & 7.6$\times 10^{-7}$& 7.7$\times 10^{-7}$& 7.2$\times 10^{-7}$& 7.7$\times 10^{-7}$& 7.5$\times 10^{-7}$& 6.5$\times 10^{-7}$\\ 
				&&$\underline{+}$4.8$\times 10^{-7}$&$\underline{+}$4.4$\times 10^{-7}$&$\underline{+}$4.4$\times 10^{-7}$&$\underline{+}$4.3$\times 10^{-7}$& $\underline{+}$4.6$\times 10^{-7}$&$\underline{+}$4.5$\times 10^{-7}$&$\underline{+}$4.0$\times 10^{-7}$\\
				\cline{2-9}
				& minimum & 1.6$\times 10^{-6}$ & 1.7$\times 10^{-6}$ & 1.7$\times 10^{-6}$& 2.0$\times 10^{-6}$& 2.2$\times 10^{-6}$& 2.1$\times 10^{-6}$& 2.2$\times 10^{-6}$\\
				&bias &$\underline{+}$8.6$\times 10^{-7}$&$\underline{+}$9.8$\times 10^{-7}$&$\underline{+}$9.7$\times 10^{-7}$&$\underline{+}$1.2$\times 10^{-6}$& $\underline{+}$1.3$\times 10^{-6}$&$\underline{+}$1.3$\times 10^{-6}$&$\underline{+}$1.3$\times 10^{-6}$\\
				\hline				
				
				& 0-20& 6.6$\times 10^{-6}$& 7.7$\times 10^{-6}$& 7.8$\times 10^{-6}$& 9.6$\times 10^{-6}$&1.1$\times 10^{-5}$&1.0$\times 10^{-5}$& 1.1$\times 10^{-5}$\\
				&&$\underline{+}$2.5$\times 10^{-6}$&$\underline{+}$3.1$\times 10^{-6}$&$\underline{+}$3.1$\times 10^{-6}$&$\underline{+}$4.1$\times 10^{-6}$& $\underline{+}$4.6$\times 10^{-6}$&$\underline{+}$4.4$\times 10^{-6}$&$\underline{+}$4.7$\times 10^{-6}$\\
				\cline{2-9}
				&20-40& 5.5$\times 10^{-6}$& 6.2$\times 10^{-6}$& 6.3$\times 10^{-6}$& 8.1$\times 10^{-6}$& 8.7$\times 10^{-6}$&8.5$\times 10^{-6}$& 9.0$\times 10^{-6}$ \\
				&&$\underline{+}$2.1$\times 10^{-6}$&$\underline{+}$2.5$\times 10^{-6}$&$\underline{+}$2.5$\times 10^{-6}$&$\underline{+}$3.5$\times 10^{-6}$& $\underline{+}$3.8$\times 10^{-6}$&$\underline{+}$3.7$\times 10^{-6}$&$\underline{+}$4.0$\times 10^{-6}$\\
				\cline{2-9}			
				17.0& 40-60 & 4.2$\times 10^{-6}$ & 4.8$\times 10^{-6}$& 4.9$\times 10^{-6}$&6.3$\times 10^{-6}$&6.8$\times 10^{-6}$&6.6$\times 10^{-6}$&6.4$\times 10^{-6}$\\
				&&$\underline{+}$1.6$\times 10^{-6}$&$\underline{+}$1.9$\times 10^{-6}$&$\underline{+}$1.9$\times 10^{-6}$&$\underline{+}$2.7$\times 10^{-6}$& $\underline{+}$2.9$\times 10^{-6}$&$\underline{+}$2.9$\times 10^{-6}$&$\underline{+}$2.8$\times 10^{-6}$\\
				\cline{2-9}
				& 60-80 & 3.1$\times 10^{-6}$ & 3.4$\times 10^{-6}$& 3.2$\times 10^{-6}$& 3.9$\times 10^{-6}$& 4.2$\times 10^{-6}$& 4.1$\times 10^{-6}$& 3.9$\times 10^{-6}$\\
				&&$\underline{+}$1.2$\times 10^{-6}$&$\underline{+}$1.4$\times 10^{-6}$&$\underline{+}$1.3$\times 10^{-6}$&$\underline{+}$1.7$\times 10^{-6}$& $\underline{+}$1.8$\times 10^{-6}$&$\underline{+}$1.8$\times 10^{-6}$&$\underline{+}$1.7$\times 10^{-6}$\\
				\cline{2-9}
				& 80-100 & 2.4$\times 10^{-6}$& 2.2$\times 10^{-6}$& 2.2$\times 10^{-6}$& 2.1$\times 10^{-6}$& 2.3$\times 10^{-6}$&2.2$\times 10^{-6}$& 1.9$\times 10^{-6}$\\ 
				&&$\underline{+}$9.0$\times 10^{-7}$&$\underline{+}$8.7$\times 10^{-7}$&$\underline{+}$8.7$\times 10^{-7}$&$\underline{+}$9.0$\times 10^{-7}$& $\underline{+}$9.8$\times 10^{-7}$&$\underline{+}$9.6$\times 10^{-7}$&$\underline{+}$8.5$\times 10^{-7}$\\
				\cline{2-9}
				& minimum & 4.4$\times 10^{-6}$&4.8$\times 10^{-6}$ & 4.9$\times 10^{-6}$& 6.0$\times 10^{-6}$& 6.4$\times 10^{-6}$& 6.3$\times 10^{-6}$& 6.4$\times 10^{-6}$\\
				&bias&$\underline{+}$1.6$\times 10^{-6}$&$\underline{+}$1.9$\times 10^{-6}$&$\underline{+}$1.9$\times 10^{-6}$&$\underline{+}$2.6$\times 10^{-6}$& $\underline{+}$2.8$\times 10^{-6}$&$\underline{+}$2.7$\times 10^{-6}$&$\underline{+}$2.8$\times 10^{-6}$\\
				\hline
			\end{tabular}
		}
		\caption{$J/\psi$ yield in different centrality classes for different p+A collision systems at different collision energies.}		
		\label{table14}
	\end{table*}
	\FloatBarrier
	
	\FloatBarrier
	\begin{table*}[h]
		\centering
		\resizebox{1.2\columnwidth}{!}
		{
			\hskip-2.0cm\begin{tabular}{|c|c|c|c|c|}
				\hline
				{ $\sqrt{s_{NN}}$ (GeV)} & { centrality (\%)}  & {$<N_{NiNi}^{J/\psi}>$}  & {$<N_{AuAu}^{J/\psi}>$}& {$<N_{PbPb}^{J/\psi}>$} \\
				\hline
				&0-20& 1.4$\times 10^{-7}$ $\underline{+}$2.1$\times 10^{-7}$ & 3.7$\times 10^{-7}$ $\underline{+}$7.5$\times 10^{-7}$ & 3.9$\times 10^{-7}$ $\underline{+}$7.8$\times 10^{-7}$ \\
				\cline{2-5}
				&20-40& 5.5$\times 10^{-8}$ $\underline{+}$8.3$\times 10^{-8}$ & 1.4$\times 10^{-7}$ $\underline{+}$2.7$\times 10^{-7}$ & 1.4$\times 10^{-7}$ $\underline{+}$2.8$\times 10^{-7}$ \\
				\cline{2-5}
				7.6&40-60& 2.0$\times 10^{-8}$ $\underline{+}$3.0$\times 10^{-8}$ & 4.0$\times 10^{-8}$ $\underline{+}$7.9$\times 10^{-8}$ & 4.1$\times 10^{-8}$ $\underline{+}$8.2$\times 10^{-8}$ \\
				\cline{2-5}
				&60-80& 6.6$\times 10^{-9}$ $\underline{+}$9.9$\times 10^{-9}$ & 1.1$\times 10^{-8}$ $\underline{+}$2.3$\times 10^{-8}$ & 5.9$\times 10^{-9}$ $\underline{+}$1.2$\times 10^{-8}$ \\
				\cline{2-5}
				&80-100& 2.2$\times 10^{-9}$ $\underline{+}$3.3$\times 10^{-9}$ & 1.1$\times 10^{-9}$ $\underline{+}$2.3$\times 10^{-9}$ & 1.2$\times 10^{-9}$ $\underline{+}$2.3$\times 10^{-9}$ \\
				\cline{2-5}
				&minimum bias& 4.4$\times 10^{-8}$ $\underline{+}$6.6$\times 10^{-8}$ & 1.1$\times 10^{-7}$ $\underline{+}$2.3$\times 10^{-7}$ & 1.2$\times 10^{-7}$ $\underline{+}$2.3$\times 10^{-7}$ \\ 
				\cline{2-5}
				\hline
				&0-20& 5.3$\times 10^{-5}$ $\underline{+}$3.3$\times 10^{-5}$ & 2.0$\times 10^{-4}$ $\underline{+}$1.4$\times 10^{-4}$ & 2.0$\times 10^{-4}$ $\underline{+}$1.4$\times 10^{-4}$ \\
				\cline{2-5}
				&20-40& 5.3$\times 10^{-5}$ $\underline{+}$1.3$\times 10^{-5}$ & 2.0$\times 10^{-4}$ $\underline{+}$5.1$\times 10^{-5}$ & 2.0$\times 10^{-4}$ $\underline{+}$5.1$\times 10^{-5}$ \\
				\cline{2-5}
				12.3&40-60& 2.1$\times 10^{-5}$ $\underline{+}$4.8$\times 10^{-6}$ & 7.2$\times 10^{-5}$ $\underline{+}$1.5$\times 10^{-5}$ & 7.2$\times 10^{-5}$ $\underline{+}$1.5$\times 10^{-5}$ \\
				\cline{2-5}
				&60-80& 7.5$\times 10^{-6}$ $\underline{+}$1.6$\times 10^{-6}$ & 2.1$\times 10^{-5}$ $\underline{+}$4.2$\times 10^{-6}$ & 2.1$\times 10^{-5}$ $\underline{+}$2.1$\times 10^{-6}$ \\
				\cline{2-5}
				&80-100& 8.4$\times 10^{-7}$ $\underline{+}$5.3$\times 10^{-7}$ & 6.0$\times 10^{-7}$ $\underline{+}$4.2$\times 10^{-7}$ & 6.0$\times 10^{-7}$ $\underline{+}$4.3$\times 10^{-7}$ \\
				\cline{2-5}
				&minimum bias& 1.7$\times 10^{-5}$ $\underline{+}$1.1$\times 10^{-5}$  & 6.0$\times 10^{-5}$ $\underline{+}$4.2$\times 10^{-5}$ & 6.0$\times 10^{-5}$ $\underline{+}$4.3$\times 10^{-5}$ \\ 
				\cline{2-5}			
				\hline
				&0-20& 1.6$\times 10^{-4}$ $\underline{+}$7.6$\times 10^{-5}$ & 6.4$\times 10^{-4}$ $\underline{+}$3.5$\times 10^{-4}$ & 6.6$\times 10^{-4}$ $\underline{+}$3.6$\times 10^{-4}$ \\
				\cline{2-5}
				&20-40& 6.4$\times 10^{-5}$ $\underline{+}$3.0$\times 10^{-5}$ & 2.3$\times 10^{-4}$  $\underline{+}$1.3$\times 10^{-4}$  & 2.4$\times 10^{-4}$ $\underline{+}$1.3$\times 10^{-4}$ \\
				\cline{2-5}
				17.0&40-60& 2.3$\times 10^{-5}$ $\underline{+}$1.1$\times 10^{-5}$ & 6.8$\times 10^{-5}$ $\underline{+}$3.7$\times 10^{-5}$ & 6.9$\times 10^{-5}$  $\underline{+}$3.8$\times 10^{-5}$ \\ 
				\cline{2-5}
				&60-80& 7.7$\times 10^{-6}$ $\underline{+}$3.6$\times 10^{-6}$ & 1.9$\times 10^{-5}$ $\underline{+}$1.1$\times 10^{-5}$ & 9.8$\times 10^{-6}$ $\underline{+}$5.4$\times 10^{-6}$ \\
				\cline{2-5}
				&80-100& 2.6$\times 10^{-6}$ $\underline{+}$1.2$\times 10^{-6}$  & 1.9$\times 10^{-6}$  $\underline{+}$1.1$\times 10^{-6}$ & 2.0$\times 10^{-6}$ $\underline{+}$1.1$\times 10^{-6}$ \\
				\cline{2-5}
				&minimum bias& 5.1$\times 10^{-5}$ $\underline{+}$2.4$\times 10^{-5}$ & 1.9$\times 10^{-4}$ $\underline{+}$1.1$\times 10^{-4}$ & 2.0$\times 10^{-4}$ $\underline{+}$1.1$\times 10^{-4}$ \\ 
				\cline{2-5}			
				\hline
			\end{tabular}
		}
		\caption{$J/\psi$ yield in different centrality classes for different A+A collision systems at different collision energies.}		
		\label{table15}
	\end{table*}
	\FloatBarrier
	
	\section{Summary}
	This article presents a data driven approach to estimate the $J/\psi$ production cross-sections for p+A and A+A collision systems in the energy domain that will be explored at future experimental facilities studying charm production in relatively low energy nuclear collisions ($\sqrt{s}\sim$~7.6-17.0~GeV). Data available from different fixed target p+A collisions in the SPS energy domain have been utilized for this purpose. The geometrical Glauber model has been employed to estimate the influence of cold nuclear medium on $J/\psi$ production, via an effective absorption cross section $\sigma_{abs}^{J/\psi}$, which shows a significant energy dependence, with increased dissociation at lower collision energies. Phenomenological parametrizations of the existing experimental data in p+p and p+A collisions are finally used to calculate the production rates at lower energies for different collision systems, energies and centralities. For low energy heavy-ion collisions our goal is to estimate the production in absence of any secondary medium, thus providing a reference benchmark with which the experimental data can be contrasted as and when available. Our estimated yield values for these foreseen collision systems will be useful for the upcoming experiments to perform realistic physics performance simulations and thus optimize their detection capabilities. 
	
	\section*{Acknowledgment}
	The authors are grateful to Z. Xu for useful discussions. S. Chatterjee would like to thank Dr. S. Biswas, Bose Institute, Kolkata for his support during the course of the study. He further acknowledges the receipt of the institutional fellowship of Bose Institute from DST, Government of India.

\end{document}